\documentclass{emulateapj}

\usepackage{enumerate}

\def\Msun{M$_{\odot}$}

\def\Mstar{$M_{\star}$}

\def\ha{${\rm H}\alpha$}
\def\hb{${\rm H}\beta$}
\def\oiii{[O\,{\scriptsize III}]}
\def\oii{[O\,{\scriptsize II}]}
\def\nii{[N\,{\scriptsize II}]}
\def\lya{${\rm Ly}\alpha$}

\shorttitle{\ha\ mapping on the USS~1558-003 proto-cluster}
\shortauthors{Hayashi M. et al.}

\begin{document}


\title{A star-bursting proto-cluster in making associated
to a radio galaxy at $z=2.53$ discovered by H$\alpha$ imaging}


\author{Masao Hayashi\altaffilmark{1}, Tadayuki
  Kodama\altaffilmark{1,2}, Ken-ichi Tadaki\altaffilmark{1,3}, Yusei
  Koyama\altaffilmark{1,4}, and Ichi Tanaka\altaffilmark{2}} 



\altaffiltext{1}{Optical and Infrared Astronomy Division, National Astronomical Observatory, Mitaka, Tokyo 181-8588, Japan; masao.hayashi@nao.ac.jp}
\altaffiltext{2}{Subaru Telescope, National Astronomical Observatory of Japan, 650 North A'ohoku Place, Hilo, HI 96720, USA}
\altaffiltext{3}{Department of Astronomy, Graduate School of Science, University of Tokyo, Tokyo 113-0033, Japan}
\altaffiltext{4}{Department of Physics, Durham University, South Road, Durham DH1 3LE}


\begin{abstract}
We report a discovery of a proto-cluster in vigorous assembly and
hosting strong star forming activities, associated to a radio galaxy
USS~1558-003 at $z$=2.53, as traced by a wide-field narrow-band \ha\
imaging with MOIRCS on Subaru Telescope. We find 68 \ha\ emitters with
dust-uncorrected SFRs down to 8.6 \Msun\ yr$^{-1}$. Their spatial
distribution indicates that there are three prominent clumps of \ha\
emitters, one surrounding the radio galaxy and another located at
$\sim$1.5 Mpc away to the south-west, and the other located in
between the two. These contiguous three systems are very likely to
merge together in the near future and may grow to a single more
massive cluster at later times. Whilst most \ha\ emitters reside 
in the ``blue cloud'' on the color--magnitude diagram, some emitters
have very red colors with $J-K_s>1.38$(AB). Interestingly, such red
\ha\ emitters are located towards the faint end of the red sequence,
and they tend to be located in the high density clumps. We do not see
any statistically significant difference in the distributions of
individual star formation rates or stellar masses of the \ha\ emitters
between the dense clumps and the other regions, suggesting that this
is one of the notable sites where the progenitors of massive galaxies
in the present-day clusters were in their vigorous formation
phase. Finally, we find that \ha\ emission of the radio galaxy is
fairly extended spatially over $\sim4.5$\arcsec. However it is not as
widespread as its \lya\ halo, meaning that the \lya\ emission is
indeed severely extended by resonant scattering.  
\end{abstract}





\section{Introduction}
It is well-known that galaxy properties are strongly dependent on the
environment where galaxies reside \citep[e.g.,][]{dressler1997,tanaka2005,cucciati2006}.  
Galaxy clusters are one of the most biased environments in the
Universe, and local ones are dominated by galaxies with red colors
and elliptical morphologies which have already quenched
their star forming activities and are passively evolving. 
In contrast, star forming galaxies with a blue color and
spiral/irregular morphologies are preferentially found in the field
environment. This trend is known as a star formation -- density
relation where star formation activity of galaxies decreases gradually
in higher density regions \citep[e.g.,][]{kauffmann2004,cooper2008}. 
It is certain that some processes specific to high density environment
play an important role in formation and evolution of elliptical
galaxies found in the local Universe. 
The plausible environmental effects include
galaxy--intracluster medium (ICM) interactions such as ram-pressure stripping
\citep{Gunn1972}, galaxy--cluster gravitational interactions
\citep{Byrd1990}, and galaxy--galaxy interactions such as mergers
and harassment \citep{Moore1996}.
However, the relative importance of the processes
still remains unclear.   

The fraction of blue star forming galaxies in galaxy clusters
increases with redshifts up to $z\sim1$ 
\citep[Butcher-Oemler effect,][]{butcher1978,butcher1984}.    
Recent observations have revealed that there are many starburst
galaxies even in high density regions at $z\gtrsim 1.5$ 
\citep{hayashi2010,hilton2010,papovich2010,tran2010,fassbender2011}. 
However, as their color--magnitude diagrams show a prominent red
sequence, a large fraction of passive galaxies also exist in these
clusters, and thus formation epoch of such massive galaxies seems to
be much earlier. Indeed, \citet{gobat2011} have found that there are
already mature galaxies in a cluster at $z=2.07$ which is the most
distant $X$-ray cluster to date.
It is essential to survey dense regions at higher redshifts of
$z\gtrsim2$ in more detail to investigate the site where a large
fraction of progenitors of local elliptical are evolving vigorously
and reveal critical processes for their evolution, although these
recent studies obviously suggest that star forming activity of
galaxies become more active with increasing redshifts. 
The importance of surveys in high density regions at $z\gtrsim2$ is
supported by the fact that the activities of galaxies and AGNs have
peaks at $z=$1--3 \citep[i.e.,][]{madau1996,hopkins2006,ueda2003}. 
Moreover, it should be noted that massive galaxies on the red sequence
disappear in proto-clusters at $z\sim3$ \citep{kodama2007}, while such
red galaxies are already in place in proto-clusters at $z\sim2$
\citep[see also][]{kajisawa2006,kriek2008,doherty2010}.   

Proto-clusters at $z\gtrsim2$ are ideal targets to investigate the
environmental dependence of galaxy properties at high
redshift. High-$z$ radio galaxies (HzRGs) are used as good landmarks to
search for proto-clusters, because they are thought to be progenitors of
massive elliptical galaxies located in the center of local galaxy clusters
\citep[e.g.,][]{McLure1999}. 
Indeed, some over-density regions around HzRGs of various galaxy populations such
as Lyman $\alpha$ emitters (LAEs), Lyman break galaxies (LBGs),
distant red galaxies (DRGs), are identified as proto-clusters
\citep[e.g.,][]{pentericci1997,Miley2004,kurk2004a,kurk2004b,kajisawa2006,kodama2007,Kuiper2010}. 
Therefore narrow-band imaging surveys targeting \ha\ emitters (HAEs)
around the HzRG
is also an effective method to investigate star formation activity in
a high-$z$ proto-cluster.  

There are already several surveys of \ha\ emission lines in 
proto-clusters at $z\gtrsim2$; PKS~1338-262 proto-cluster at $z=2.16$
\citep{kurk2004a,hatch2011}, 4C+10.48 proto-cluster at $z=2.35$
\citep{hatch2011}, 4C23.56 proto-cluster at $z=2.48$
\citep{tanaka.I2011}. \citet{hatch2011} have found statistical excesses 
in number density of \ha\ emitting galaxies in the vicinity of the
radio galaxies of PKS~1338-262 and 4C+10.48 by a factor of $>$ 10 compared
with that of general fields. The \ha\ luminosity functions in the proto-clusters
have similar shapes to those in the general fields, but the normalization of
the luminosity function in the proto-clusters is a factor of 13 higher than
that in the fields.
\citet{tanaka.I2011} have found a statistical excess in the number density of
\ha\ emitters in the 4C23.56 proto-cluster by a factor of 5 compared to a
general field.
Combining with mid-infrared photometric data which traces reradiation by dust,
they revealed that active star formation must been occurring in the proto-clusters,
which is as active as that in the general fields at similar redshifts.
They also suggest that it is probable that star formation activity in
proto-clusters gets stronger towards higher redshifts.  
Therefore, it is interesting to probe star formation activity in higher redshift
proto-clusters to confirm this trend.

\begin{figure}
\epsscale{1.0}
\plotone{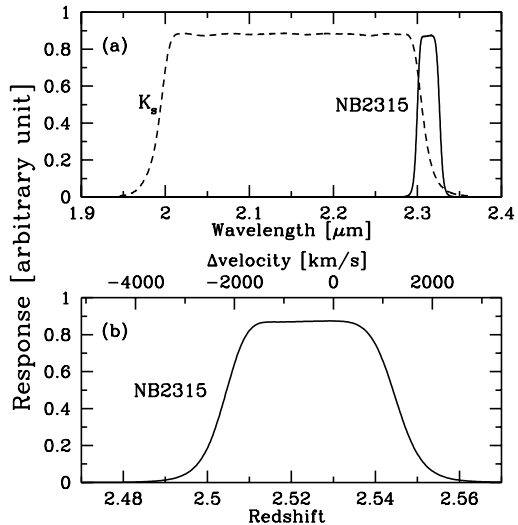}
\caption{(a) The broken line shows a response curve of MOIRCS $K_s$
  broad-band filter, and the solid line shows that of MOIRCS NB2315
  narrow-band filter ($\lambda_c=2.313\micron, \Delta\lambda=0.027\micron$). 
 (b) The response curve of NB2315 is shown as a function of redshift.
  The relative line-of-sight velocity with respect to the radio galaxy
  redshift ($z=2.53$) is also shown for the case of \ha\ emission line.
\label{fig;filter}}
\end{figure}

In this paper, we present results of our \ha\ emission line survey in
the 4\arcmin$\times$7\arcmin\ region around a radio galaxy
USS~1558-003 at $z=2.53$ with the NB2315 narrow-band filter
($\lambda_c=2.313\micron, \Delta\lambda=0.027\micron$). This filter is
designed to perfectly match to this particular proto-cluster (Figure
\ref{fig;filter}). The redshift of 2.53 is the highest where \ha\
emission lines can be traced by near-infrared imaging on a
ground-based telescope. \citet{kajisawa2006} and \citet{kodama2007}
have reported a statistical excess of bright DRGs around this radio
galaxy as an evidence for the existence of a proto-cluster associated
to the radio galaxy at $z=2.53$. In the current paper, our studies
have been conducted under the MAHALO-Subaru project (MApping H-Alpha
and Lines of Oxygen with Subaru; Kodama et al., in prep). This project
aims to reveal the environmental dependence of star forming activities
by wide-field narrow-band imaging of \ha\ or \oii\ emission lines in
(proto-)clusters and general fields at $1.5 \lesssim z \lesssim 2.5$. 
Refer also to Kodama et al.~(in prep) for the details, such as the
method of selection of emission line galaxies and the estimation of
galaxy properties (star formation rates and stellar masses).
Our previous studies have demonstrated that the narrow-band imaging is
very powerful and useful to completely sample star forming galaxies
with emission lines down to a certain limiting flux
\citep{kodama2004,koyama2010,koyama2011,hayashi2010,hayashi2011,tadaki2011,tanaka.I2011}.

The structure of this paper is the following.
Observations and available data are described in \S~\ref{sec;data}. 
Our sample of \ha\ emitters at $z=2.53$ around the proto-cluster are
selected from the photometric catalog in \S~\ref{sec;HAE}.  
We investigate the spatial distribution and colors of the \ha\
emitters in \S~\ref{sec;results}.  Discussions on star forming
activity, properties of \ha\ emitters on the red sequence, and the
\ha\ emission of the radio galaxy will follow in
\S~\ref{sec;discussions}. Finally, we summarize our results of this
paper in \S~\ref{sec;conclusions}.     
Throughout this paper, magnitudes are presented in the AB system, and
we adopt cosmological parameters of $h=0.7$, $\Omega_{m}=0.3$ and
$\Omega_{\Lambda}=0.7$. Vega magnitudes in $J$, $H$ and $K_s$, if
preferred, can be obtained from our AB magnitudes using the following
calibrations: $J$(Vega)=$J$(AB)$-$0.94, $H$(Vega)=$H$(AB)$-$1.38, and
$K$(Vega)=$K$(AB)$-$1.86, respectively. At $z=2.53$, 1 arcmin
corresponds to 0.483 Mpc (physical) and 1.705 Mpc (comoving),
respectively.

\section{Observation and Data}
\label{sec;data}

\begin{deluxetable*}{ccccccc}
\tabletypesize{\scriptsize}
\tablecaption{Summary of the optical and near-infrared images.}
\tablewidth{0pt}
\tablehead{
\colhead{Filter} & \colhead{FoV\tablenotemark{a}} &
\colhead{Integration time} & \colhead{Limiting mag.\tablenotemark{b}} &
\colhead{Seeing} & \colhead{Instrument} & \colhead{Observation date} \\
\colhead{}  & \colhead{} &
\colhead{(minutes)} & \colhead{(5$\sigma$)} & 
\colhead{(arcsec)} & \colhead{} & \colhead{} 
}
\startdata
$B$    & F1+F2 & 80 & 27.16 & 0.70 & Suprime-Cam & 2011 April 29 \\
$r'$   & F1+F2 & 90 & 26.87 & 0.63 & Suprime-Cam & 2011 April 29 \\
$z'$   & F1+F2 & 55 & 25.75 & 0.66 & Suprime-Cam & 2011 April 29 \\
$J$    & F2    & 75 & 24.18 & 0.42 & MOIRCS      & 2011 March 11 \\
$H$    & F2    & 45 & 23.51 & 0.47 & MOIRCS      & 2011 March 11 \\
$K_s$  & F1+F2 & 57 & 23.65 & 0.66 & MOIRCS      & 2011 March 11, 2011 April 17 \\
       &       & (F1: 32, F2: 25) & (F1: 23.46, F2: 23.17) & (F1: 0.66, F2: 0.40) &    & \\
NB2315 & F1+F2 & 203 & 23.01 & 0.53 & MOIRCS     & 2011 March 11, 2011 April 17 \\
       &       & (F1: 133, F2: 70) &  (F1: 22.74, F2: 22.35) & (F1: 0.53, F2: 0.36) &   &  
\enddata
\tablecomments{Finally, the FWHMs of PSF in all the images are matched
  to 0.66\arcsec, except for the $B$-band image which has a FWHM of 0.70\arcsec.}
\tablenotetext{a}{The pointings of F1 and F2 for $K_s$ and NB2315 images
have an offset of 1\arcmin\ to the west and 1\arcmin\ to the south.}
\tablenotetext{b}{The limiting magnitudes are measured with
a 1.5\arcsec\ diameter aperture.}
\label{table;obsanddata}
\end{deluxetable*}

Observations of the USS~1558-003 proto-cluster were conducted on 2011
March 11, April 17 and 29 as a Subaru open-use intensive 
program (S10B-028, PI: T.~Kodama). The optical and near-infrared
(NIR) images were taken with Subaru Prime Focus Camera
\citep[Suprime-Cam;][]{miyazaki2002} and Multi-Object Infra-Red 
Camera and Spectrograph \citep[MOIRCS;][]{ichikawa2006,suzuki2008}
on Subaru telescope, respectively.
Suprime-Cam has a field-of-view (FoV) of 27$\times$34 arcmin$^2$,
while MOIRCS has a FoV of 4$\times$7 arcmin$^2$. 
The $K_s$ and NB2315 data were obtained with two pointings with an
offset of 1\arcmin\ in right ascension and 1\arcmin\ in declination,
to neatly cover the dense clumps of \ha\ emitters (see \S \ref{sec;select-ha})
which were recognized during the course of observing runs.
We call these two pointings as F1 and F2.
Although the panoramic Suprime-Cam image covers the entire region of
both F1 and F2, the MOIRCS $J$ and $H$ data are available only in the
F2 pointing.
Therefore, it should be noted that the area used in
this study is limited to the F2 region of about 4$\times$7 arcmin$^2$
where both optical and NIR data are available.
The weather was fine during the observations, and the sky condition
was photometric. The seeing was less than 0.66\arcsec\ in all the images
except for 0.70\arcsec\ in the $B$-band. The total integration times in
the optical and NIR broad-bands range from 45 to 90 minutes. We took significantly
longer integration (203 minutes) at the narrow-band NB2315.
Consequently, the available data set consists
of six broad-band data, $B,r',z'$ in optical and $J,H,K_s$ in NIR, and
the narrow-band data with the NB2315 filter. 

All of the optical and NIR data are reduced in a standard manner using 
a data reduction package for Suprime-Cam \citep[SDFRED  ver.2.0:][]{ouchi2004} 
and MOIRCS (MCSRED\footnote{\url{http://www.naoj.org/staff/ichi/MCSRED/mcsred.html}}
by I.~Tanaka), respectively. Refer to \citet{hayashi2010,hayashi2011}
for the details of the reduction procedures.
PSFs in all the reduced images, except for the $B$-band image, are matched to
0.66 arcsec. The zero-points of magnitudes are determined using the
standard stars, GD153 and LDS749B for optical data, and FS27, FS137
and G191-B2B for NIR data. The 5$\sigma$ limiting magnitudes are 27.16,
26.87, 25.75, 24.18, 23.51, 23.65, and 23.01 in $B$, $r'$, $z'$, $J$,
$H$, $K_s$, and NB2315, respectively, which are measured with a
1.5\arcsec\ diameter aperture. The details of the observation and data
are summarized in Table~\ref{table;obsanddata}.

\section{\ha\ emitters in the USS~1558-003 proto-cluster}
\label{sec;HAE}

\subsection{Photometric catalog}
\label{sec;catalog}
Source detection is performed on the original NB2315 image with a FWHM
of PSF of 0.53\arcsec rather than the PSF matched image, using {\sc SExtractor}
\citep[ver.~2.5.0:][]{bertin1996}, and photometry on all the images is
conducted by the double-image mode of {\sc SExtractor}.   
Color indices are defined with aperture magnitudes {\sc mag\_aper},
which are measured with a 1.5\arcsec-diameter aperture, while the
total magnitudes are defined with {\sc mag\_auto} magnitudes
which are measured with an elliptical aperture \citep[see also][]{Kron1980}.
Magnitude errors are estimated from 1$\sigma$ sky noise at each object
position taking account of slightly different exposure times and
sensitivities, which is thus corrected for the difference in depth
over the entire FoV. The regions near the edges of each image where
the exposure time is shorter than 75\% of the total in each combined
image are masked. If an object is not detected in a broad-band filter,
a 2$\sigma$ limiting magnitude is used to estimate the upper/lower
limits in color indices or the upper limits in magnitudes. We make
sure that the stellar colors are consistent with those of stellar
templates given in \citet{gunn1983} in order to check the zero-points
of magnitudes. From such comparison, the zero-point magnitudes in $B$,
$r'$, $z'$ and $J$ are corrected by $\sim0.15$ magnitude at most so
that the stellar colors are in good agreement with those of the
stellar atlas. Moreover, magnitudes are corrected for the Galactic
absorption by the following magnitudes; A($B$)=0.62, A($r'$)=0.41,
A($z'$)=0.23, A($J$)=0.14, A($H$)=0.09, A($K_s$)=0.06, and
A(NB912)=0.05 which are derived from the extinction law of
\citet{cardelli1989} on an assumption of $R_V=3.1$ and $E(B-V)=0.155$
based on \citet{schlegel1998}. 

As a result, 1,035 objects are detected in the NB2315 image at more
than 5$\sigma$ level (i.e., 23.01 mag.\ in the F1+F2 region). Among
them, 754 galaxies are distinguished from 281 stars based on their
$B-z'$ and $z'-K_s$ colors. This method to separate galaxies from
stars are devised by \citet{daddi2004} \citep[see also][]{kong2006}. 

The detection completeness is investigated as follows\footnote{
This completeness refers to point sources, and the
one for extended objects can be relatively shallower than that.
However, the sizes of galaxies at $z\sim2.5$ are not generally large
enough to be resolved in the current data, and
the assumption of point sources may not be so impractical.  
}.
First, artificial objects with given magnitudes and Gaussian profiles
with FWHM of 0.53\arcsec\ are distributed on the NB2315 image. Source
detection is conducted in the same manner as described above, and
then the fraction of detected artificial objects is calculated. The
detection completeness is found to be more than 74 (83) \% down to
23.0 (22.0) magnitude in NB2315.

\subsection{Selection of \ha\ emitters}
\label{sec;select-ha}

\begin{figure}
\epsscale{1.0}
\plotone{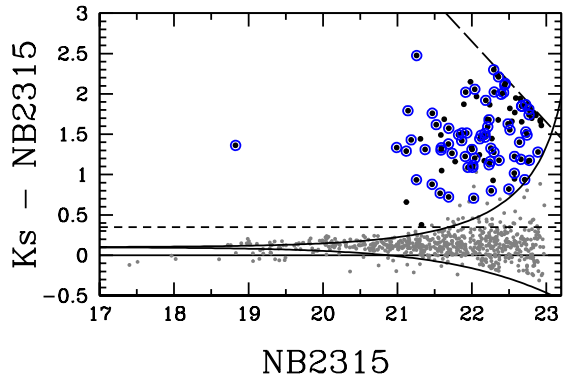}
\caption{
  The color--magnitude diagram of $K_s$$-$NB2315
  versus NB2315. The solid curves show the boundary of 3$\sigma$ excess given
  by the equation (\ref{eq;colorexcess}). The broken line shows the 2$\sigma$
  limit in $K_s$$-$NB2315 color. The black dots located above the solid
  curve indicate the NB2315 emitters. Some objects with large enough color
  excesses are not classified as emitters due to the shallow data at
  the individual object positions. Blue open circles with the black dots
  represent the \ha\ emitters associated to the radio galaxy at $z=2.53$,
  which are identified by the color selection criteria in $r'-J$ and $J-K_s$
  (equation (\ref{eq;rjk}) and Figure~\ref{fig;rjk}). 
  \label{fig;bb-nb}}
\end{figure}

\begin{figure*}
\epsscale{1.15}
\plottwo{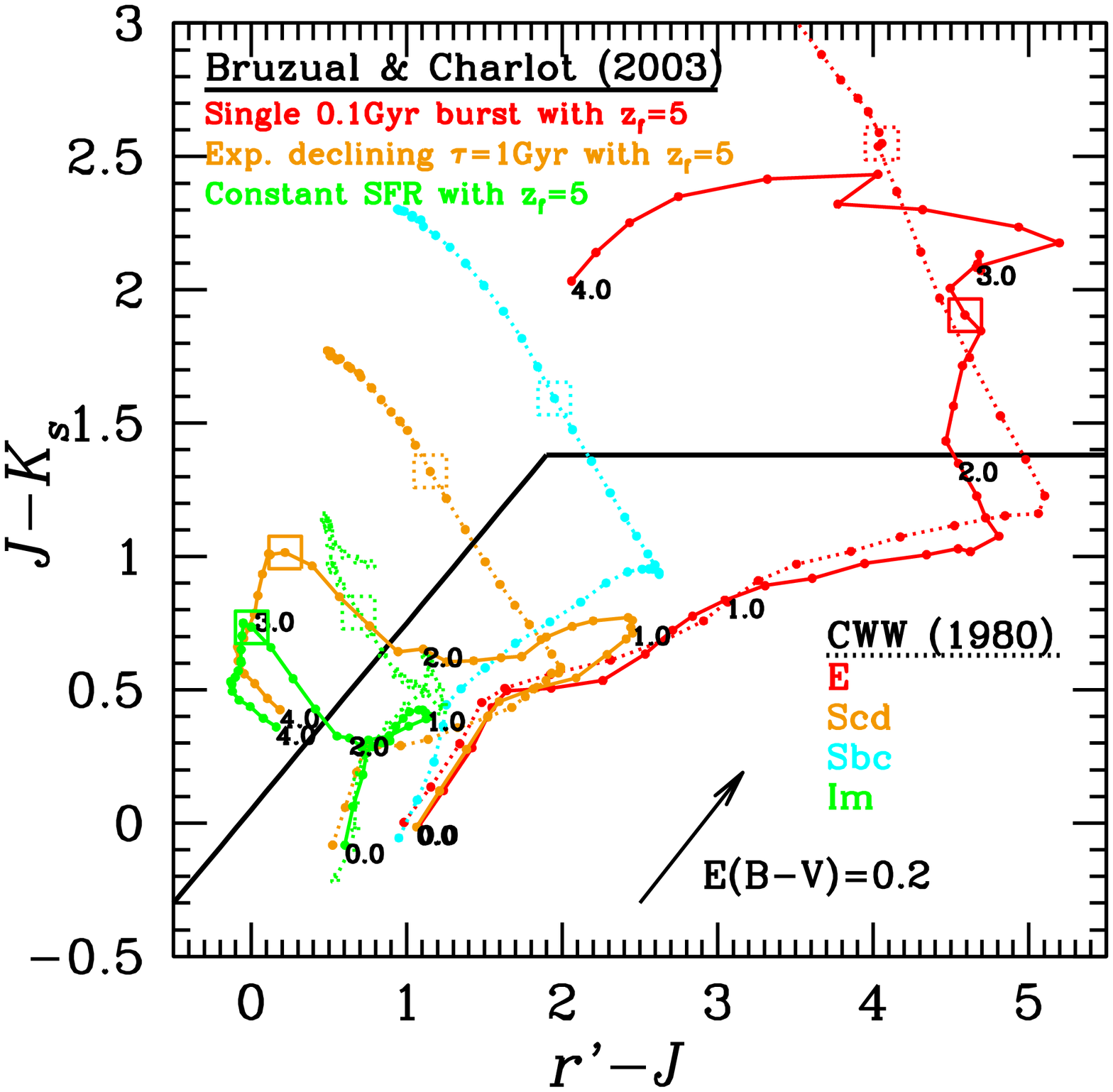}{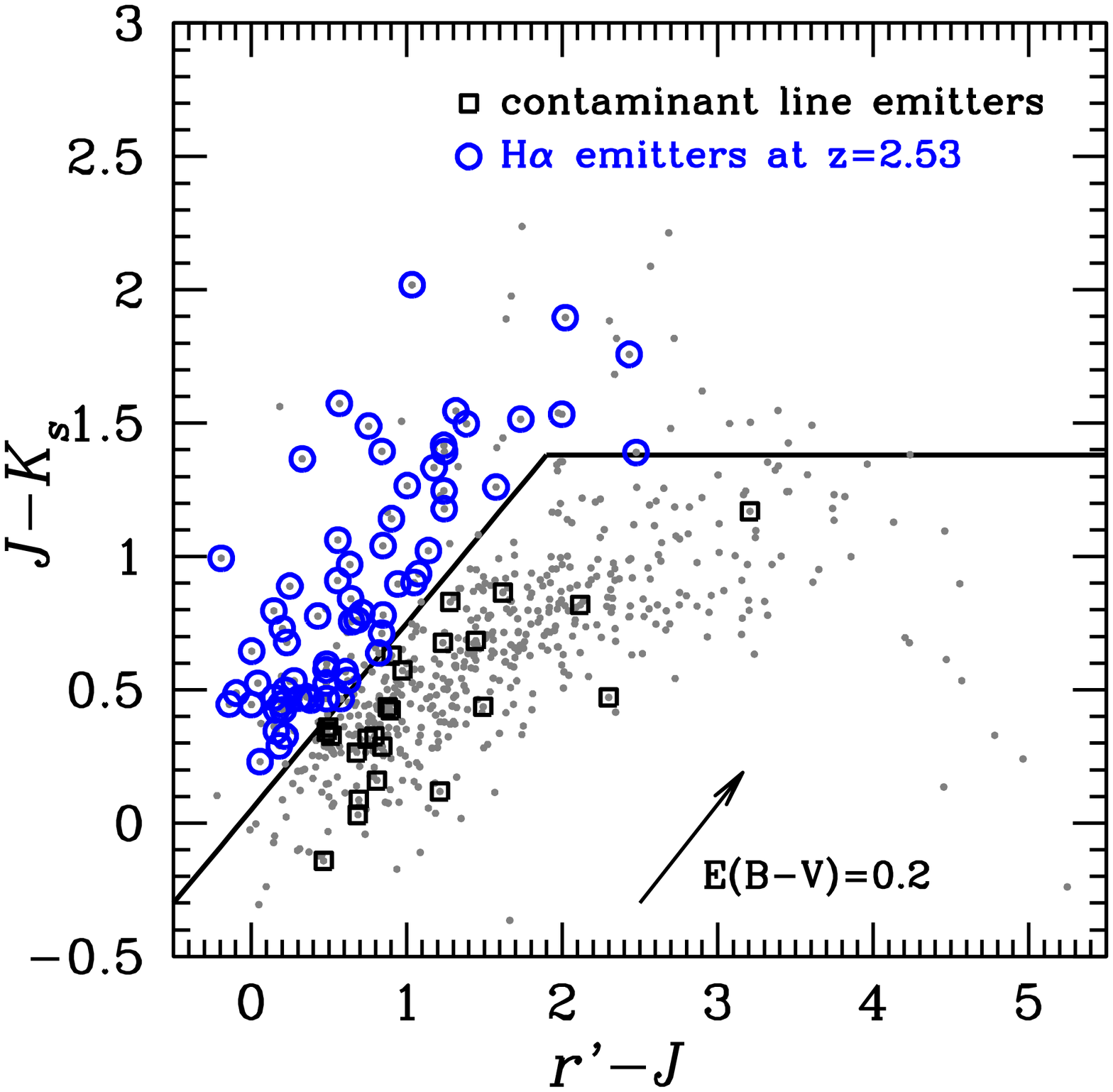}
\caption{(Left panel) The color--color diagram of $r'-J$
  versus $J-Ks$. Red, orange, and green solid lines show color tracks
  of model SEDs for three types of star formation histories with fixed
  formation redshift of $z_f=5$ based on \citet{bc03}. Dotted lines
  show the color tracks created by four empirical spectral templates from
  elliptical to irregular type galaxies compiled by \citet{coleman1980}. 
  Dust extinction is not corrected, but the absorption by neutral hydrogen
  given by \citet{madau1995} is taken into account. The arrow shows
  a reddening vector of $E(B-V)=0.2$, which is estimated from the dust
  extinction curve of \citet{calzetti2000}. The numbers shown along
  the tracks for \citet{bc03} model indicate redshifts, and the open
  squares correspond to $z=2.5$. The black folded solid line shows the
  boundaries of our selection criteria to identify \ha\ emitters among
  the NB2315 emitters (equation (\ref{eq;rjk})). (Right panel) Same as
  the left panel, but the observed galaxy colors are plotted.
  Gray dots show the NB2315-detected galaxies. Blue open circles show
  our sample of \ha\ emitters, and black open squares show the other
  contaminant line emitters probably located at lower-$z$ ($z<1$). 
\label{fig;rjk}}
\end{figure*}

Galaxies with a nebular emission line that happens to enter the NB2315 narrow-band
filter should be observed much brighter in NB2315 magnitude compared with the $K_s$
broad-band magnitude which samples primarily the underlying continuum flux density.
We thus apply the following criteria to select such galaxies with a
emission line at $\sim2.313\micron$;
\begin{equation}
K_s-NB > -2.5\log(1-\sqrt[]{\mathstrut f^2_{3\sigma,K_s}+f^2_{3\sigma,NB}}/f_{NB})+0.1,
\label{eq;colorexcess}
\end{equation}
\begin{equation}
K_s-NB > 0.35, 
\end{equation}
where $f_{3\sigma}$ is the 3$\sigma$ sky noise flux in each band and
$f_{NB}$ is the NB2315-band flux density (Figure \ref{fig;bb-nb}). The
continuum flux density is corrected for a color term by 0.1 magnitude as given
in the right most side of equation (\ref{eq;colorexcess}), because the
effective wavelength of $K_s$ filter is different by 0.163\micron\
from that of NB2315 filter (Figure \ref{fig;filter}). We note that the
$K_s$--NB2315 colors of \citet{coleman1980} templates
redshifted to $z=2.53$ are actually distributed around 0.1 magnitude. 
The second criterion of $K_s$--NB$>$0.35 corresponds to the observed
equivalent width larger than 79\AA\ (22\AA\ in the rest-frame if a galaxy
is located at $z$=2.53) after being corrected for the color term of 0.1
magnitude.
We select galaxies with color excesses in $K_s$--NB2315
greater than 3$\sigma$ photometric errors. This means that a emission line
with a flux larger than 2.6$\times$10$^{-17}$ erg s$^{-1}$ cm$^{-2}$
can be firmly detected. If the color excess is due to a \ha+\nii\ emission
line pair at $z=2.53$, the limiting line flux corresponds
to L(\ha)=$1.1\times10^{42}$ erg s$^{-1}$ and a dust-free SFR of 8.6
\Msun\ yr$^{-1}$ \citep{kennicutt1998}, where the ratio of \nii\ to
\ha (\nii/\ha) is assumed to be 0.22 \citep{sobral2012a}.  
We also find 12 galaxies whose line fluxes are larger than this
limit, but color excesses are smaller than the 3$\sigma$ level
due to their location near the edge of the images where the depth is shallower.
These galaxies are not included in our emitter sample.
As a result, we select 102 NB2315 emitters out of 754 galaxies over
the $\sim 4\times7$ arcmin$^2$ area (Figure~\ref{fig;bb-nb}).

Although we aim to select \ha\ emitters at $z=2.53$, it is possible
that some of the emission lines detected with the NB2315 filter are
some other lines at different redshifts such as \hb\ or \oiii\ at
$z\sim3.6$, \oii\ at $z=5.2$, and Paschen series lines at
$z<1.0$. However, it is unlikely that the bulk of the NB2315 emitters
originate from \hb, \oiii, or \oii\ lines at $z>3.5$, because the
color excesses of the emitters are quite large as shown in Figure \ref{fig;bb-nb}.  
Note that the large color excesses we see in $K_s$--NB2315 may correspond to
unrealistically large emission line fluxes, if they are all located at such
high redshifts.
Indeed, all but seven NB2315 emitters are detected in $B$-band and 64
out of 68 \ha\ emitters classified below also have the detection.   
If a nebular line detected by the NB2315 filter would be either an
\oiii\ or an \oii\ line, the line luminosity of \oiii\ (or \oii) is
larger than $2.0(4.7)\times10^{42}$ erg s$^{-1}$, respectively.
The luminosity functions of \oiii\ at $z$=0.84 \citep{ly2007} and
\oii\ at $z=1.47$ \citep{sobral2012a} imply that no more than one
\oiii\ or \oii\ emitter with such a high luminosity can be included in
our NB2315 emitter sample with limited field coverage, if any.
Therefore, we believe that there is few contamination from the lines
at $z>3.5$ in our NB2315 emitter sample. Moreover, it should be noted
that our NB2315 survey is targeting the plausible proto-cluster at
$z$=2.53 associated to the radio galaxy, and we must be preferentially
detecting \ha\ emitters at $z=2.53$ rather than other high-$z$
emitters in the general field. However, as reported by
\citet{geach2008} who conducted the H$_2$S1 narrow-band
($\lambda_c=2.121$\micron) imaging with WFCAM on UKIRT to search for
\ha\ emitters at $z=2.23$ in the COSMOS field, other lower-$z$
emission lines such as Pa$\alpha$ and Pa$\beta$ at $z<1$ can be
significant contaminations. In order to discriminate \ha\ from other
lower-$z$ contaminant lines, we set the following color
selection criteria on the color--color diagram, $r'-J$ versus $J-K_s$;   
\begin{equation}
\begin{tabular}{lcl}
$J-K_s > 0.7(r'-J)+0.05$ & for & $r'-J < 1.9$ \\
$J-K_s > 1.38$           & for & $r'-J > 1.9$.
\end{tabular}
\label{eq;rjk}
\end{equation}
This classification with $r'JK_s$ colors is an analogue of the $BzK$ color
selection of galaxies at $1.4 \lesssim z \lesssim 2.5$ devised by
\citet{daddi2004}, but it is optimized to identify galaxies at
$z \gtrsim 2.5$. Figure \ref{fig;rjk} shows the color tracks of model
spectra with various star formation histories and fixed formation
redshift of $z_f=5$ based on \citet{bc03}, and four SED templates 
in \citet{coleman1980} which are redshifted from $z=0.0$ to $z=4.0$.
This figure suggests that the criteria enable us not
only to select DRGs as galaxies with red colors ($J-K_s$(vega)$>2.3$),
but also to isolate star forming galaxies at $z \gtrsim 2.5$ from
those at lower redshifts. Although no dust extinction is taken into
account for the color tracks, the arrow in the figure shows the
reddening vector of $E(B-V)=0.2$ based on the \citet{calzetti2000}
extinction curve. This indicates that our selection criteria are relatively
free from the dust reddening effect, as the vector is almost parallel to
the slanted boundary line. 
Indeed, a group of emission line galaxies which are likely to be \ha\
emitters at $z=2.53$ seem to be located separately from the bulk of
galaxies at lower redshifts as shown in the right panel of Figure
\ref{fig;rjk}. Thus, our selection criteria work well to select \ha\
emitters at $z=2.53$. 

We attempted to apply other color selection methods such as the original
$Bz'K_s$ selection \citep{daddi2004} and $JHK_s$ selection as well
\citep{kajisawa2006,kodama2007}.    
However, the $Bz'K_s$ selection is effective to select galaxies
at $1.4 \la z \la 2.5$, while we aim to pick out \ha\ emitters at $z\sim2.5$
among the NB2315 emitters.
Since they are located just at the upper edge of the redshift range
of the $Bz'K$ selection, the selection of \ha\ emitters may well be
quite incomplete.
On the other hand, it is found that the $JHK_s$ selection do not work
well due to the shallowness of $H$-band data and the incompleteness
in the selection of blue star-forming galaxies.
Therefore, we judge that the classification with $r'JK_s$
colors works best among them to identify \ha\ emitters at $z=2.53$.
Instead of the color selection, one may consider the use of
photometric redshifts to discriminate among different emission lines
at different redshifts.
We also derive photometric redshifts of our NB2315-detected
galaxies using six bands photometry of $B,r',z',J,H,K_s$ and the EAZY code
\citep{brammer2008}. However, we find that it is difficult to cleanly
discriminate \ha\ emitters from other possible lines, especially those
at $z>1.0$, based on the photometric redshifts.
In the end, we decide to use the $r'JK_s$ diagram to identify
\ha\ emitters. Consequently, we select 68 \ha\ emitters in total
(Figure~\ref{fig;rjk}).

\subsection{Selection of Distant Red Galaxies}
\label{sec;select-DRG}

\begin{figure}
\epsscale{1.1}
\plotone{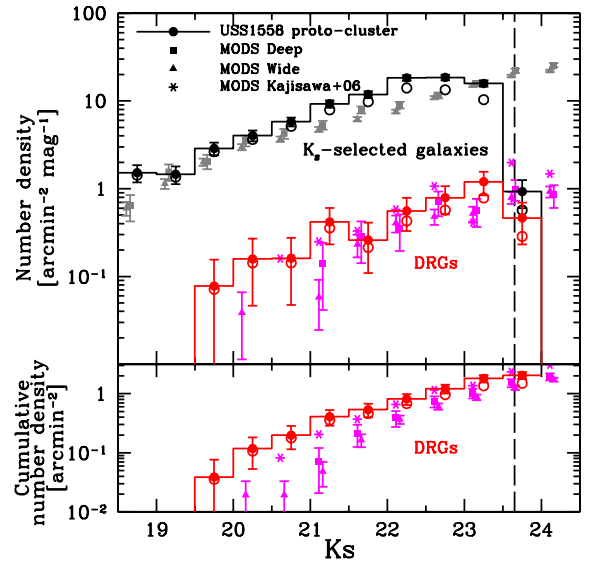}
\caption{Differential number densities of DRGs and $K_s$-selected
  galaxies around the USS~1558-003 proto-cluster are shown in upper
  panel, and cumulative number densities of DRGs are shown in lower
  panel. Filled circles show the number densities corrected for
  detection incompleteness, while open circles show those without
  corrections. Filled squares and triangles show the results obtained
  by MODS deep and wide surveys \citep{Kajisawa2011}, and asterisks
  show those of \citet{Kajisawa2006PASJ}. The results of MODS deep and
  wide are not corrected for detection incompleteness. Error bars are
  estimated based on Poisson statistics. Vertical broken line shows
  the 5$\sigma$ limiting magnitude in $K_s$.    
  \label{fig;nd_drg}}
\end{figure}

DRG is a class of galaxy populations at $2\la z \la 4$ with red colors of
$J-K_s>1.38$ or $J-K_s$(vega)$>2.3$ \citep{Franx2003} which consist of
dusty starburst galaxies and quiescent galaxies. 
To select such red galaxies in this region, the DRG color selection is
applied to the $K_s$-selected catalog which is made with the same procedures as those 
described in \S~\ref{sec;catalog} except that source detection is performed
on the $K_s$ image. The $K_s$-selected catalog is used only for the selection
of DRGs, since such catalog enables us to select DRGs more completely
down to fainter magnitude than the NB2315-limiting magnitude. The catalog
includes 1,340 objects brighter than 5$\sigma$ limiting magnitude in
$K_s$. Among them, 1,002 objects are classified as galaxies, while the
remaining 338 objects are classified as stars.
Consequently, 42 DRGs are selected from the $K_s$-selected catalog.
It is expected that red galaxies associated to the proto-cluster at
$z=2.53$ tend to be seen as DRGs, although DRGs in general can be located
in a much wider redshift range of $2\la z \la 4$.

Figure \ref{fig;nd_drg} shows the number densities of DRGs and
$K_s$-selected galaxies which are corrected for detection
incompleteness estimated in the same manner as in
\S\ref{sec;catalog}. For comparison with other studies, we also plot
the results of the general blank field survey, MOIRCS Deep Survey (MODS)
\citep{Kajisawa2006PASJ,Kajisawa2011} 
where the deep and wide surveys have FoVs of 28.2 and 103.3 arcmin$^2$,
respectively.
The number densities of DRGs as well as $K_s$-selected galaxies in the
USS1558 proto-cluster do not show strong overdensities compared to
those in the MODS, although some excesses by a factor of a few are
seen at several bins. 
\citet{kodama2007} found that this region shows an overdensity
of bright DRGs compared to the GOODS-S field \citep{Giavalisco2004}.
Figure \ref{fig;nd_drg} indeed shows that there is an excess of  
bright DRGs with $K_s \la 21.5$ in the USS~1558-003 proto-cluster
which are almost absent in the general blank field.
The difference in number densities of DRGs between the proto-cluster
and the general field is clearer in the cumulative densities,
showing the excess at $\sim 2 \sigma$ level in the bin with $K_s<21.5$.  
Moreover, the spatial distribution of DRGs is not homogeneous
but clustered around the radio galaxy (see \S\ref{sec;map}).
The fact that there is no significant excess of faint DRGs in the
proto-cluster region may suggest that most of the faint member
galaxies in this proto-cluster have bluer colors due probably to
on-going star formation activities, and that passive red galaxies
would be gradually emerging later on.
Although we do not have yet any spectroscopic confirmation of membership
of the DRGs except for the radio galaxy itself, it is likely that
the region around the USS~1558-003 radio galaxy is a proto-cluster
at $z=2.53$. 

\begin{deluxetable}{cccccc}
\tabletypesize{\scriptsize}
\tablecaption{The number and number density of \ha\ emitters (HAEs) and DRGs.}
\tablewidth{0pt}
\tablehead{
\colhead{Region} & \colhead{Area} & \multicolumn{2}{c}{Number} & \multicolumn{2}{c}{Density} \\
 & \colhead{(arcmin$^{2}$)} &  &  &  \multicolumn{2}{c}{(arcmin$^{-2}$)}\\
\cline{3-4} \cline{5-6}\\[-2mm]
 & & HAE & DRG & HAE & DRG
}
\startdata
Clump 1    &  3.36 &  15 &  12 &  4.46$\pm$1.15 &  3.57$\pm$1.03\\
Clump 2    &  1.64 &  20 &   8 & 12.2$\pm$2.73 &  4.88$\pm$1.72\\
Clump 3    &  0.94 &   8 &   3 &  8.51$\pm$3.01 &  3.19$\pm$1.84\\
All clumps &  5.94 &  43 &  23 &  7.24$\pm$1.10 &  3.87$\pm$0.81\\
Others     & 21.16 &  25 &  19 &  1.18$\pm$0.24 &  0.90$\pm$0.21\\
Entire field  & 27.10 &  68 &  42 &  2.51$\pm$0.30 &  1.55$\pm$0.24
\enddata
\tablecomments{Errors in the number density are estimated based on
  Poisson statistics.}
\label{table;numHAEDRG}
\end{deluxetable}

\section{Results}
\label{sec;results}

\subsection{Spatial distribution of \ha\ emitters}
\label{sec;map}

\begin{figure}
\epsscale{1.3}
\plotone{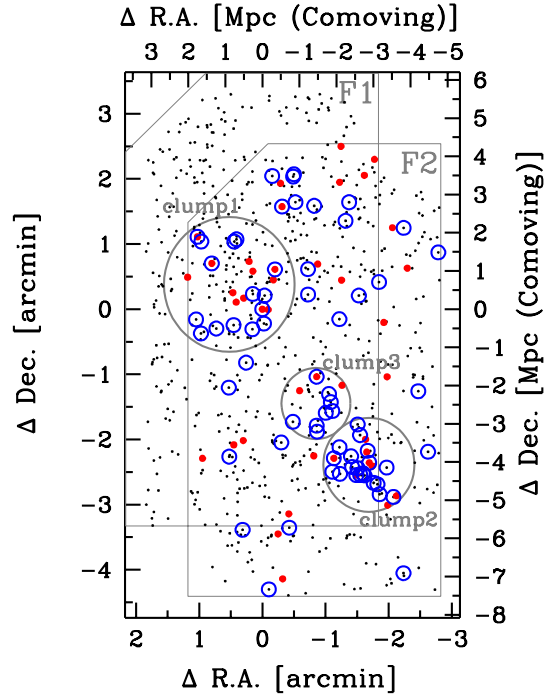}
\caption{The spatial distribution of \ha\ emitters. Blue open circles
  show \ha\ emitters at $z=2.53$. Red filled circles shows DRGs, and
  black dots are NB2315-detected galaxies. North is up, and east is to
  the left. The origin of the coordinates is the position of the
  USS~1558-003 radio galaxy. Three gray circles are the regions where
  \ha\ emitters and DRGs are strongly clustered, which are defined as
  clump-1, clump-2 and clump-3, respectively. Two regions enclosed by
  gray solid lines show our MOIRCS pointings (F1 and F2). 
  \label{fig;map_HAE}}
\end{figure}

Figure \ref{fig;map_HAE} shows the spatial distribution of
our 68 \ha\ emitter (HAE) candidates associated to the USS~1558-003
radio galaxy at $z=2.53$ located at (0,0) in the figure.
The 42 DRGs are also plotted with red filled circles.
We find that there are three outstanding regions where HAEs
and/or DRGs are strongly clustered, which we hereafter call clump-1,
clump-2, and clump-3. In fact, as summarized in Table
\ref{table;numHAEDRG}, the number densities of HAEs and DRGs in these
three clumps are all higher than those of averaged values across the
observed fields by factors of 2--5 for HAEs and 2--3 for DRGs. The
clump-1 is the vicinity of the radio galaxy, and it contains both HAEs
and DRGs around the radio galaxy. If we assume that the central
dominant radio galaxy grows to a cD galaxy in the future, this region
may correspond to the central part of the cluster. The fact that DRGs
are also clustered in this clump may also suggest that this part of
the proto-cluster is the oldest.
The clump-2 is the most conspicuous, densest association of the HAEs
and DRGs, and is located at $\sim3.2$\arcmin\ (about 1.5 Mpc in
physical scale) away from the radio galaxy to the south-west. The
clump-3 is a smaller group of \ha\ emitters located in between the
clump-1 and the clump-2.

It is interesting to note that the \ha\ emitters are more strongly
clustered towards the south-west clumps (clump-2 and clump-3)
rather than in the immediate surrounding region around the radio
galaxy (clump-1). They constitute a part of large scale structure
hosting the radio galaxy and clump-1, and they would all merge together
in the near future to form a more massive single cluster around the
radio galaxy. It is obvious that this proto-cluster region is not
relaxed yet and just in the process of galaxy assembly from the the
surrounding regions. We note that such distribution of the \ha\
emitters is similar to that of DRGs reported by \citet{kodama2007}
based on NTT/SOFI imaging data. In fact, we also confirm that the DRGs
tend to be located in and along the structures traced by the \ha\
emitters based on our deeper MOIRCS data (Figure \ref{fig;map_HAE}).
Furthermore, we find that these DRGs also meet the $BzK$ criteria
which select galaxies primarily at $1.4 \la z \la 2.5$, suggesting
that significant fraction of these DRGs are likely to be physically
associated to the proto-cluster hosting the radio galaxy at $z=2.53$. 

The three clumps in the 1558-003 proto-cluster host a large number of
\ha\ emitters. The surface number density of \ha\ emitters in the
three clumps is $\sim$37$\pm$13 times larger than
those in the redshift slice at $z=2.2$ in the GOODS-North field
\citep[\ha\ emission survey with NB209 narrow-band filter;][]{tadaki2011}.   
Similarly, we also find that it seems that the surface number density
is $\sim$14$\pm$3 and 17$\pm$4 times larger than those in the redshift
slices of $z=$2.2 and 2.53 in the SXDS field (\ha\ emission survey
with NB209 and NB2315; Tadaki et al.~in prep.).     

\citet{tanaka.I2011} conducted a \ha\ emitters survey in the field
around the 4C~23.56 radio galaxy at $z=2.48$ with CO narrow-band filter
on MOIRCS, and found 11 \ha\ emitters candidates to flux down
to $\sim7.5\times10^{-17}$ erg~s$^{-1}$~cm$^{-1}$ and rest-frame
EW $>$ 50\AA\ over a 23.6 arcmin$^2$ area, which is similar to our survey area.
It is found that \ha\ emitters are distributed on the east side of
the radio galaxy 4C~23.56, and there is a clump of \ha\ emitters
$\sim2$ Mpc (in comoving scale) away from the radio galaxy.
Such an offset distribution of \ha\ emitters from the radio galaxy is
similar to that in our USS~1558-003 field, although our \ha\
emitters are more strongly clustered than the 4C~23.56 field.
However, note that the EW cut and the limiting flux used for \ha\ emitter
selection are slightly different between our survey and \citet{tanaka.I2011}.
Our survey enables us to sample \ha\ emitters with fainter line fluxes and
smaller EWs. If we apply the same EW cut and the limiting flux
as in \citet{tanaka.I2011} to our sample in the USS~1558-003 field,
the number of \ha\ emitters reduces to 27, but it is still considerably
larger than that in the 4C~23.56 field.
The discovery of clumps of \ha\ emitters in these proto-clusters
around the radio galaxy at $z\sim2.5$ clearly indicates that we are
witnessing the process of mass assembly of clusters at their early stage
when galaxies are vigorously assembling to form dense cluster cores while
they are actively forming stars.

Moreover, \citet{hatch2011} report the studies for two proto-clusters
around the radio galaxies MRC~1138-262 at $z=2.16$ and 4C+10.48 at
$z=2.35$. These studies suggest that proto-clusters at $z\gtrsim2$
tend to show the statistical excess of \ha\ emitters compared to the
general fields at similar redshifts. However, they found that \ha\
emitters in the 4C+10.48 region are not strongly clustered, while
those in the MRC~1138-262 region are clustered \citep[see][]{kurk2004a}.
Moreover, both proto-clusters show inhomogeneous spatial distribution
of \ha\ emitters around each radio galaxy, and most of the \ha\
emitters are distributed only on one side of the radio galaxies,
similar to 4C~23.56 \citep{tanaka.I2011} and USS~1558-003 (this
study). 

\begin{figure}
\epsscale{1.18}
\plotone{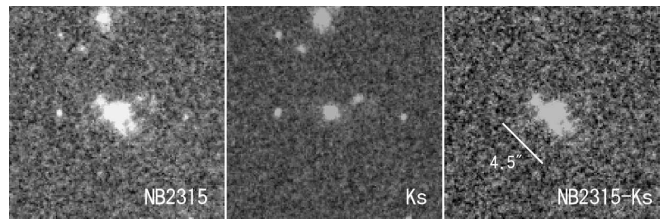}
\caption{The close-up views of the USS~1558-003 radio galaxy at $z=2.53$
  in the NB2315, $K_s$ and NB2315--$K_s$ images. Note that the FWHM
  values of PSF in these $Ks$ and NB2315 images are 0.40\arcsec and
  0.36\arcsec, respectively, since these images are created by
  combining only the frames taken under excellent seeing conditions
  $\sim0.4$\arcsec. The length of the bar in the NB2315--$K_s$ image
  indicates the angular size of 4.5\arcsec, which corresponds to 36.2
  kpc in physical scale. The NB2315--$K_s$ image clearly indicates
  that the radio galaxy has an extremely extended \ha\ emission.   
  \label{fig;haRG}}
\end{figure}

Another interesting result we find is that the radio galaxy show an
extremely extended \ha\ emission spatially as shown in Figure~\ref{fig;haRG}.
In the NB2315 image, we can notice that the structure of the \ha\
emission is stretched in the northeast--southwest direction.
The size of the \ha\ emission is $\sim4.5$\arcsec which corresponds to
$\sim$36 kpc in physical scale.  
We discuss the extended \ha\ emission of radio galaxy itself in
\S\ref{sec;radiogalaxy}. 

\subsection{Color--magnitude diagram}
\label{sec;CMD}

Color--magnitude diagram is a powerful tool to investigate the
properties of galaxies in clusters or proto-clusters. It is well-known
that galaxy clusters at low redshifts are dominated by quiescent
galaxies which make up the red sequence on the color--magnitude diagram.
The tight sequence of red galaxies is one of the prominent features seen
in galaxy clusters. However, when and how do such red quiescent galaxies
form and then evolve?  The answer to this question still remains
unclear, although there are some evidences suggesting that the blue
star forming galaxies in the early Universe evolve and become red
quiescent galaxies in high density regions during the redshift
interval of 2--3 \citep[e.g.,][]{kajisawa2006,kodama2007,kriek2008,doherty2010,gobat2011}. 
To address this, it is essential to investigate further the
color--magnitude diagram for galaxy cluster and proto-cluster at high
redshifts when the clusters are vigorously evolving.  

\begin{figure}
\epsscale{1.0}
\plotone{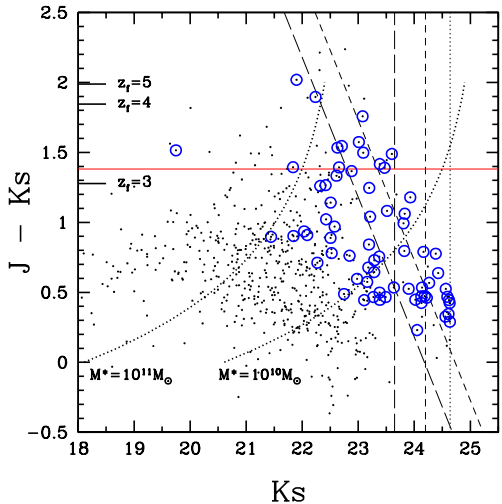}
\caption{Color--magnitude diagram of $J-K_s$ versus $Ks$. Blue open
  circles show \ha\ emitters at $z=2.53$ and black dots show all the
  galaxies in the observed field.
  Long-dashed, short-dashed, and dotted lines show 5$\sigma$,
  3$\sigma$, and 2$\sigma$ limits in color and magnitude.
  Red solid horizontal line shows the color corresponding
  to $J-K_s$=1.38, which is a criterion for selecting DRGs. The
  $J-K_s$ colors of the red quiescent galaxies with formation
  redshifts of $z_f$=3, 4, and 5 are shown by the tickmarks at the
  left edge, which are estimated by the \citet{kodama1998} model.  
  Dotted curves show the iso-stellar mass curves for
  1$\times 10^{11}$\Msun and 1$\times 10^{10}$\Msun, respectively.
  \label{fig;CMD}}
\end{figure}

\begin{figure*}
\epsscale{1.0}
\plottwo{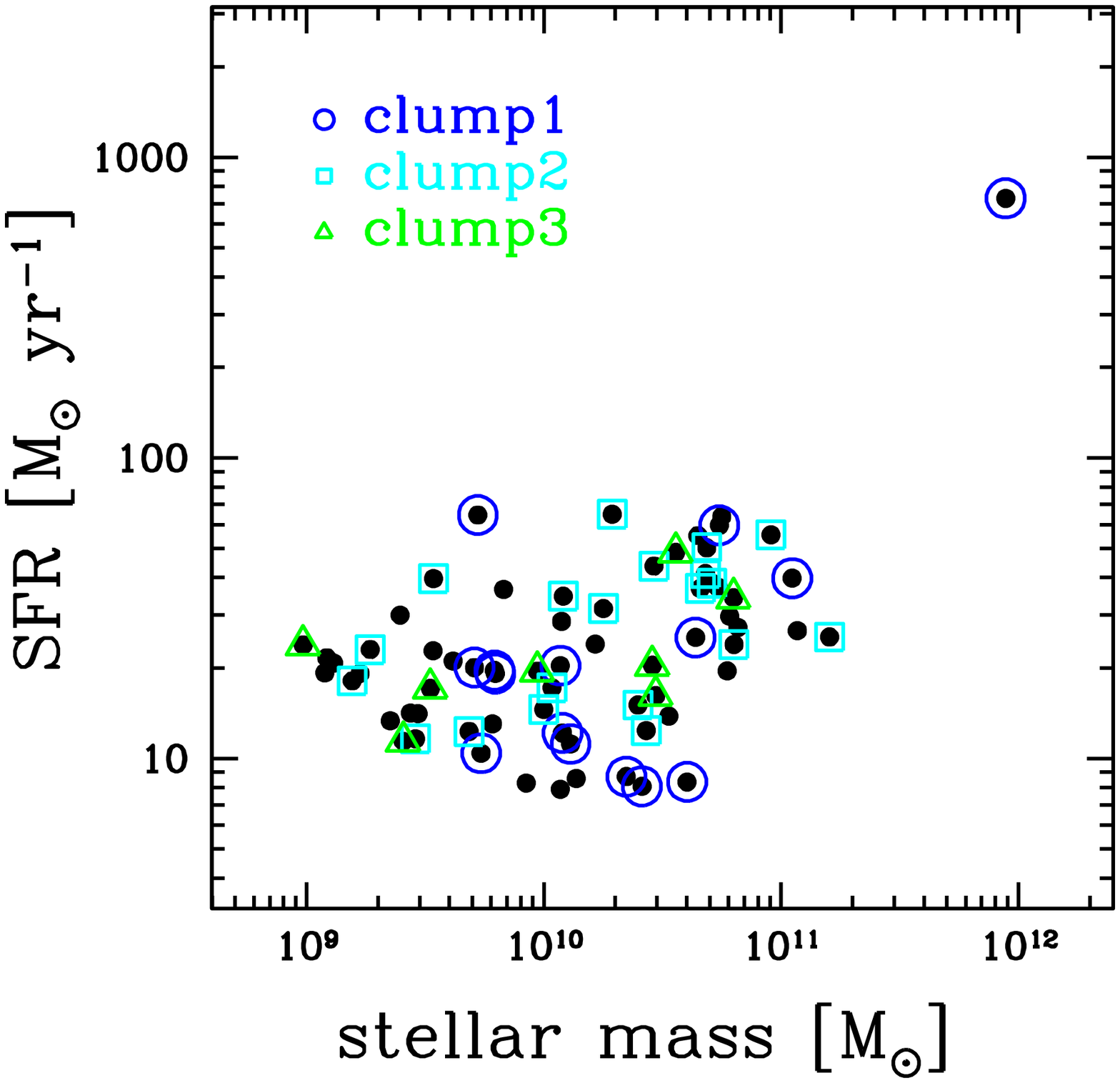}{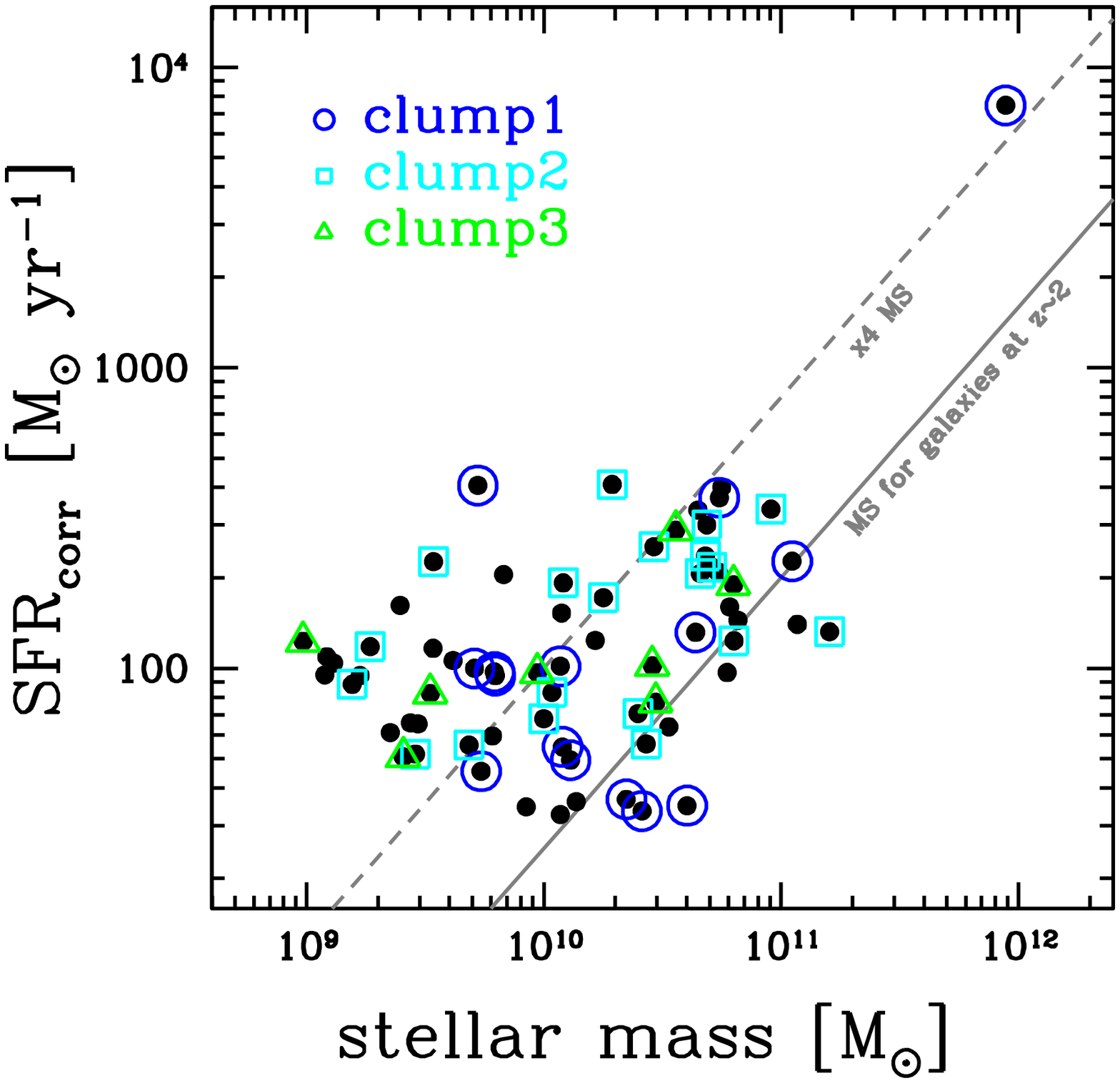}
\caption{(Left panel) Dust extinction un-corrected SFRs of the 68 \ha\
 emitters as a function of stellar mass. Those in clumps-1, 2, and 3
 are marked by blue open circles, cyan open squares, and green open
 triangles, respectively. 
 (Right panel) Same as the left panel, but the SFRs are corrected for
 dust extinction (see text for details). The solid line shows a main
 sequence (MS) of star forming galaxies at $z\sim2$
 \citep{Daddi2007,Rodighiero2011}, and the broken line corresponds to
 the SFRs elevated by a factor of 4 compared to the MS for a given
 stellar mass. 
\label{fig;sfr}}
\end{figure*}

Figure \ref{fig;CMD} shows a color--magnitude diagram of $J-K_s$
versus $K_s$ for the galaxies in the observed field towards the
proto-cluster USS~1558-003.
The \ha\ emitters associated to the proto-cluster are plotted by open circles.
As presented by \citet{kodama2007}, we may be able to recognize a sequence
of DRGs at $J-K_s\sim1.5$ which corresponds to the color of quiescent
galaxies with formation redshift of $3 \lesssim z_f < 4$ \citep{kodama1998},
as indicated by the tickmarks in the figure.
It is not surprising that the majority of the \ha\ emitters are located
on the bluer side of the diagram. However, it is intriguing that some
\ha\ emitters have very red colors satisfying the DRG criterion and constitute
mainly the faint end of the red sequence.
We also note that there is a significant color scatter among the red
\ha\ emitters.
Similarly red \ha\ emitters and dusty star forming galaxies have also been
recognized by recent observations of clusters at lower redshifts
\citep{geach2006,koyama2010,koyama2011}.
They tend to be dusty star forming galaxies and are considered to be
in the transition phase from active star forming galaxies to passive
quiescent galaxies probably under the influence of some environmental
effects, because these are preferentially found in the medium density regions
of galaxy clusters at intermediate redshifts ($z\lesssim0.8$)
when and where we see a sharp transition in the distribution
of galaxy colors from blue to red \citep{tanaka2005,koyama2008}.
Therefore it is suggestive that the faint end of the red sequence is
just being built by those transitional galaxies recognized as the red
\ha\ emitters.

For a galaxy at $z=2.53$, its $K_s$ band luminosity is still a good proxy
for the stellar mass. However, the mass-to-luminosity ratio
(\Mstar/$L_{K_s}$) is dependent on SED, and it should be corrected to
get more precise stellar mass from the $K_s$-band luminosity.
We use a simple method to estimate stellar masses of the \ha\ emitters
based on their $K_s$-band magnitudes and $J-K_s$ colors.
The relationship between $J-K_s$ colors and \Mstar/$L_{K_s}$ are
approximately estimated using the stellar population synthesis model 
of \citet{kodama1998,Kodama1999} where we constructed a sequence of
model with varying bulge-to-disk ratios. 
We note here that because of the SED degeneracy among age, metallicity,
and dust extinction, the relationship between $J-K_s$ and
\Mstar/$L_{K_s}$ is relatively insensitive to the detailed modeling of
stellar populations except for the effect of IMF variation.
We here assume the \citet{salpeter1955} IMF, and the stellar masses
in the models are scaled accordingly.
From this experiment, we establish the following relationship between
stellar mass (\Mstar), $K_s$ magnitude, and $J-K_s$ color;
\begin{equation}
\log_{10}(M_\star/10^{11}M_\odot) = -0.4(K_s-K_{11}) + \Delta \log_{10} M,
\label{eq;stellarmass}
\end{equation}
where $K_{11}$ is the $K_s$-band total magnitude corresponding to the
stellar mass of 10$^{11}$\Msun\ for a passively evolving galaxy formed
at $z$=5 and observed at $z$=2.53, which is estimated to 22.41. 
The mass-to-luminosity ratio is corrected for depending on the
$J-K_s$ color by, 
\begin{equation}
\Delta \log_{10} M = 0.12 - 1.84\cdot\exp(-1.35\cdot(J - K_s)).
\end{equation}
The iso-stellar mass curves of 1$\times 10^{11}$\Msun\ and 1$\times
10^{10}$\Msun\ are shown in Figure \ref{fig;CMD}.
All the \ha\ emitters but the radio galaxy have stellar masses of
$\lesssim 10^{11}$\Msun. Among them, the bluest \ha\ emitters with
$J-K_s<0.7$ are less massive galaxies with stellar masses of 
$\lesssim 10^{10}$\Msun. 
This gives a dichotomy in the distribution of star forming galaxies 
on the stellar mass--color plane, separated at $\sim 10^{10}$\Msun,
below which the emitters are the bluest hence the youngest or the
most actively star forming.

It should be also noted that the brightest \ha\ emitter is the radio galaxy,
and it is located at the brightest end of the red sequence.
Note that similar results are also found in other proto-clusters at $z=$2--2.5
\citep{tanaka.I2011,hatch2011}.  
Its stellar mass is estimated to $\sim10^{12}$\Msun,
which is typical for HzRGs 
\citep[e.g.,][]{Rocca-Volmerange2004,Seymour2007,Hatch2009}.   
Since the radio galaxy hosts an AGN, it is not straightforward
to quantify its star formation rate from the emission line strength
as it is affected by photoionization by AGN (\S \ref{sec;radiogalaxy}).

\subsection{Star formation activity}
\label{sec;sf}

\begin{figure}
\epsscale{1.0}
\plotone{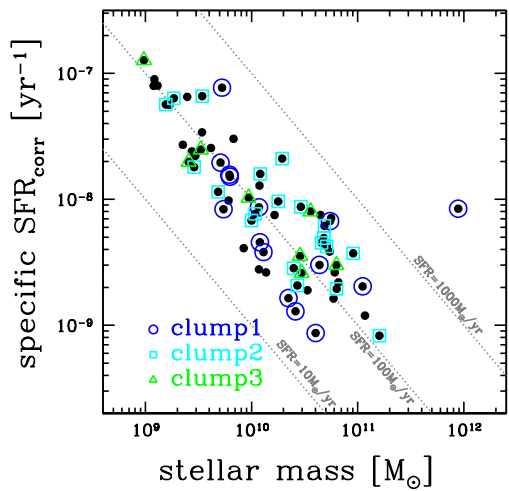}
\caption{Specific SFRs of the \ha\ emitters as a function of stellar
  mass. Dust extinction is corrected (see text for details).
  Symbols are the same as in Figure \ref{fig;sfr}.
  Three dotted lines correspond to the constant SFRs of 10, 100,
  and 1000 \Msun\ yr$^{-1}$, respectively.
\label{fig;s-sfr}}
\end{figure}

\begin{figure*}
\epsscale{1.0}
\plottwo{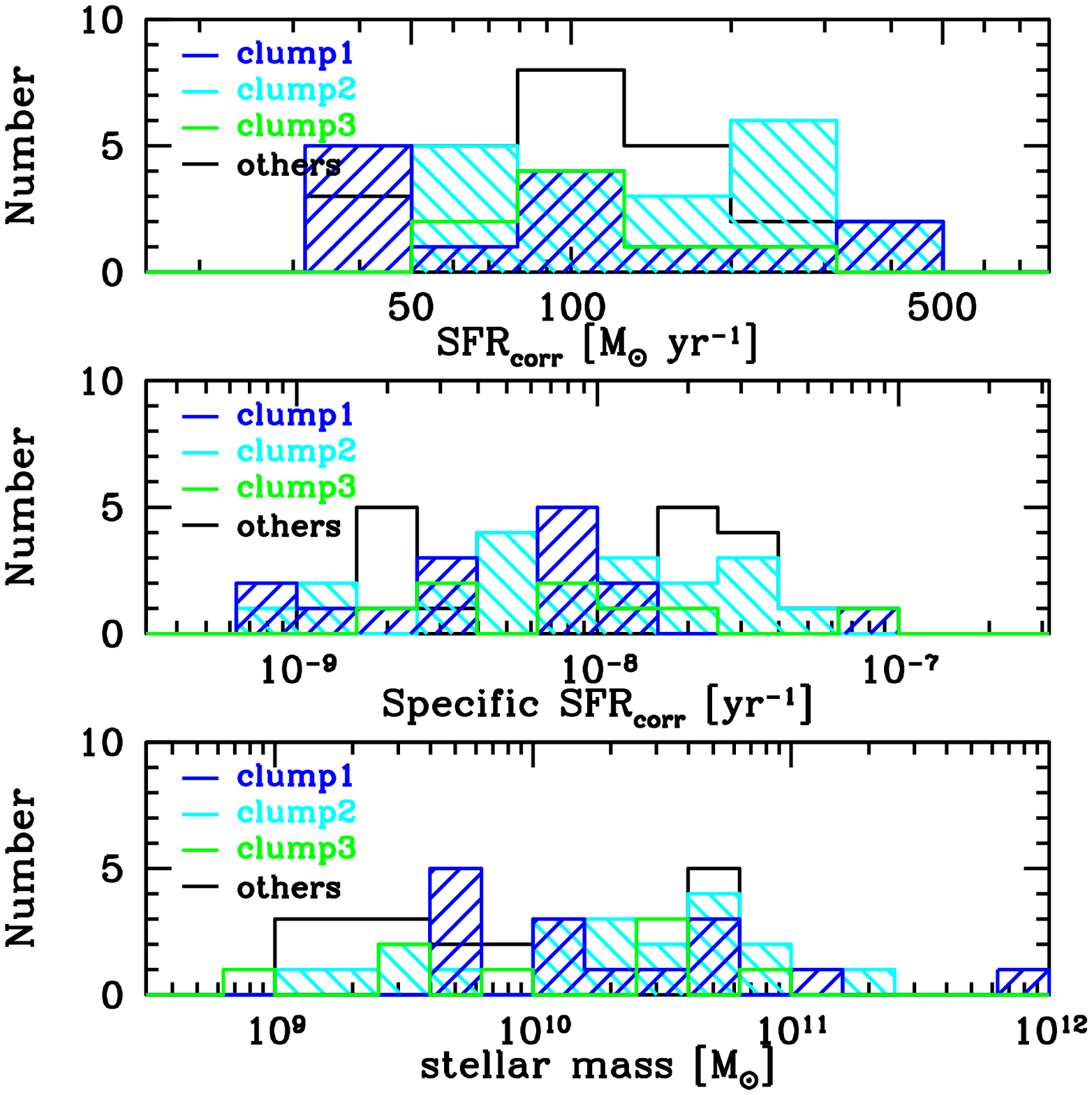}{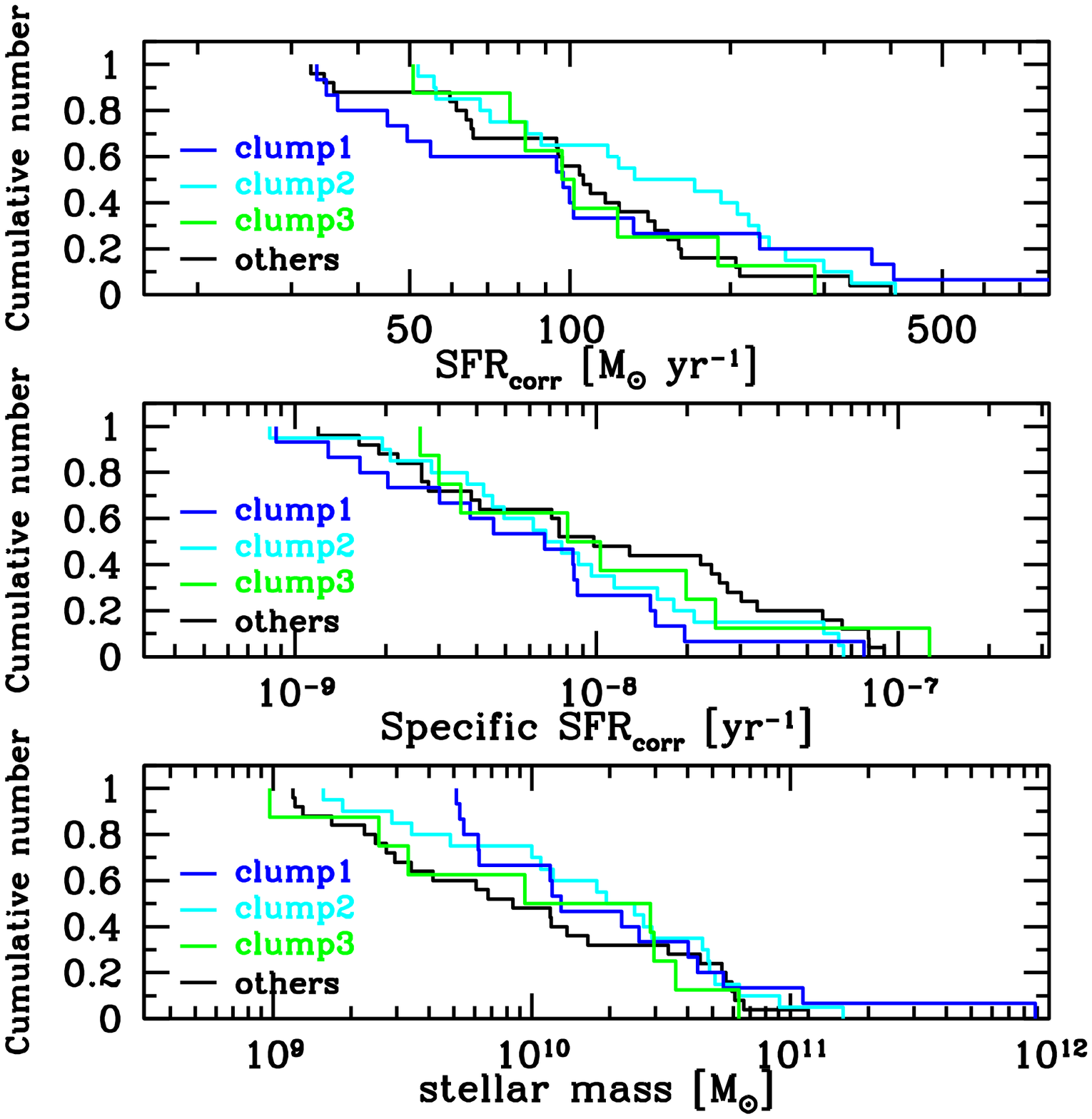}
\caption{(Left panels) Distribution of SFR (top), specific SFR (middle) and stellar
  mass (bottom) for the \ha\ emitters in the clump-1 (blue), clump-2
  (cyan),  clump-3 (green), and the other region (black),
  respectively.  
  (Right panels) Normalized cumulative number counts of the \ha\
  emitters in each region as a function of SFR (top), specific SFR
  (middle) and stellar mass (bottom).  
\label{fig;histsfr}}
\end{figure*}

The \ha\ luminosity is a good indicator of star formation rate in the
individual galaxy, because \ha\ emission line is originated from the
photoionization by young O and B stars, and thus it probes the
instantaneous star formation activity on a time scale of $<$ 20 Myr.
The advantage of using \ha\ line to estimate star formation rate
is that it is well-calibrated and widely used in the studies of
galaxies both in the local and high-$z$ Universe. Moreover
it is relatively insensitive to dust extinction compared to
\oii\ and UV luminosities.

The \ha\ luminosity is estimated from the emission line flux entering
into the narrow-band NB2315, which is the combination of \ha\ and
\nii\ lines in most cases. Since the emission lines do not enter the
broad-band filter for the \ha\ emitters at $z=2.53$,
as shown in Figure~\ref{fig;filter}, the flux
densities in the narrow-band and the broad-band filters are expressed as
$f_{NB}=f_{\rm continuum}+F({\rm H\alpha + [N\,{\scriptstyle II}]})/\Delta_{NB}$
and $f_{BB}=f_{\rm continuum}$, and thus the emission line flux is:
\begin{equation}
F({\rm H\alpha + [N\,{\scriptstyle II}]})=(f_{NB}-f_{BB})\cdot \Delta_{NB},
\label{eq;fline}
\end{equation}
where $\Delta_{NB}=271$\AA\ is FWHMs of the filters, and 
$f_{\rm continuum}$ is the flux density of continuum level of
a spectrum. It should be noted that the fluxes are derived under the
assumption that the emission lines are shifted into the center of the
narrow-band filter. Since the response curve of the filter is not
perfectly top-hat (Figure~\ref{fig;filter}), the fluxes can be
underestimated if the emission lines are not located in the center of
the filter. In that case, the SFR derived from the flux can be a lower
limit.  
The contribution of \nii\ is removed from the line flux,
F(\ha+\nii) by assuming the relation between the ratio of
\nii/\ha\ and the rest-frame equivalent width of EW$_0$(\ha+\nii)
given by \citet{sobral2012a}.

We derive SFR from \ha\ flux using the \citet{kennicutt1998} relation,
where \ha\ flux is corrected for dust extinction using the SFR-dependent
calibration given by \citet{Garn2010}. The dust extinction correction
is important even for \ha\ line which is located in the rest-frame
optical wavelength ($\lambda6563$). It is often assumed that the
amount of dust extinction, A(\ha), is unity in many papers. However,
we adopt the \citet{Garn2010} relation to more realistically estimate
the intrinsic \ha\ luminosity. The specific SFR is also calculated by
dividing the SFR by the stellar mass estimated in \S \ref{sec;CMD}.

Figure \ref{fig;sfr} shows SFRs of the \ha\ emitters as a function
of stellar mass, where dust-uncorrected and dust-corrected SFRs
(SFR$_{\rm corr}$) are shown in separate panels. There is a
correlation that more massive galaxies tend to have higher SFRs, but
the dependence is weak because there is a significant scatter in SFR
for a given stellar mass. A large fraction of the \ha\ emitters have
high SFR$_{\rm corr}$ larger than 100 \Msun\ yr$^{-1}$, meaning that many star
forming galaxies in this proto-cluster are in the starburst phase.
The object with the stellar mass of $\sim10^{12}$ \Msun\ is the radio
galaxy USS~1558-003 and it apparently shows a considerably high SFR.
However, this value must be wrong because its \ha\ line intensity
is severely contaminated by the AGN component.
In the right panel of Figure \ref{fig;sfr}, we plot the main sequence
of star forming galaxies at $z\sim2$ \citep{Daddi2007}.
In \citet{Rodighiero2011}, the galaxies whose SFRs are more than 4 times
larger than those on the main sequence for a given stellar are defined as
star-burst galaxies. Following this definition, a large fraction of
galaxies with $<10^{10}$ \Msun\ are starburst galaxies, while the
massive galaxies are more or less located around the main sequence.
This is somewhat expected by the fact that less massive galaxies
show the bluest colors on the color--magnitude diagram (Figure~\ref{fig;CMD}).
In contrast, none of massive galaxies shows such bluest colors. 
It should be noted, however, that the red \ha\ emitters with stellar masses
larger than $10^{10}$ \Msun\ are subject to large amount of dust attenuation,
and thus they are actually in dusty starburst phase as we will
discuss later in \S~\ref{sec;redHAE}. Figure \ref{fig;s-sfr} shows
specific SFRs of the \ha\ emitters plotted as a function of stellar
mass. It indicates that specific SFR and stellar mass are anti-correlated as
seen in many other studies. Massive galaxies have lower star formation
rates for a given stellar mass compared to less massive galaxies,
indicating that star forming activities have been somewhat truncated
and weakened in massive galaxies. 

In Figures \ref{fig;sfr} and \ref{fig;s-sfr}, \ha\ emitters in the
three clumps are marked with open circles, squares, and triangles.
It appears that the \ha\ emitters in the clump-2 (i.e., the south-west clump
$\sim1.5$ Mpc away from the radio galaxy) tend to have higher star formation
rates than those in other regions. To investigate any environmental
dependence in the properties of \ha\ emitters more quantitatively, we
show in Figure \ref{fig;histsfr} the differential histograms of SFRs,
specific SFRs, and stellar masses of the \ha\ emitters divided by the
regions in the left panels, and their cumulative and normalized number
counts in the right panels. 

There is no large difference in the distribution of all these physical
quantities among different regions in general, except for the weak
tendency that SFRs in the clump-2 may be slightly higher. In fact, the
median SFR of HAE in the clump-2 is 152 \Msun\ yr$^{-1}$ which
compared to 97 \Msun\ yr$^{-1}$ in the clump-1, 99 \Msun\ yr$^{-1}$ in
the clump-3 and 106 \Msun\ yr$^{-1}$ in the other regions.  
However, the Kolmogorov-Smirnov (KS) test that is applied on the top
right panel suggests that the difference does not show high enough
significance in statistics. The probability to accept the hypothesis
that the SFRs in these clumps have different distributions is less
than 2$\sigma$ at most.  
Therefore we may conclude that star formation activity at this high
redshift ($\sim$2.5) is high everywhere regardless of environment, not
only in the low density outskirts but also in the dense clumps of the 
proto-cluster. Because we are targeting a proto-cluster and its
surrounding region, it is yet unknown whether such absence of
environmental dependence at $z\sim2.5$ is universal from a
proto-cluster to another, as well as in the general low density fields
which are not in the vicinity of proto-clusters.

Furthermore, it is interesting to know how much star formation
activity and stellar mass are confined in this proto-cluster region as
a whole, as most of the structures and the \ha\ emitters are likely to
assemble to form a single richer cluster at later times. The integrated
stellar mass and SFR of the \ha\ emitters in each clump and those in
the entire region are shown in Table~\ref{table;masssfr}. 
The uncertainty in the integrated stellar mass and SFR
originated from photometric errors are small enough not to give a
large influence on the discussions below.
Also, in this analysis, we exclude the most massive radio galaxy in the clump-1
($\sim$1$\times$10$^{12}$ \Msun), since its \ha\ line flux is severely
contaminated by the AGN component. The integrated stellar mass of DRGs
associated to the proto-cluster is derived by subtracting the
component of DRGs outside the cluster but projected along the line of
slight. 
Here we use the number count of DRGs in the MODS general field survey
\citep{Kajisawa2006PASJ,Kajisawa2011}. 
We note that the total stellar masses estimated here should be taken
as the lower limits, because we miss less massive galaxies fainter
than our observational limits. 
All clumps but clump-3 have a total stellar mass of 
$\sim10^{12}$\Msun, while clump-3 has a factor of $\sim$10 lower
mass of $\sim10^{11}$\Msun.  
The integrated specific SFR of the clump-1 is slightly lower than
those of the clump-2 and the clump-3, by a factor of 1.6 and 3.9, respectively, 
which might suggest that the star formation activity per unit stellar
mass is slightly lower in the closest vicinity of the radio galaxy
which may grow to the center of the cluster in the future, and the age
of the system is slightly older than the other clumps. 
However, the statistics is again too poor for us to say anything
conclusive yet. 

\begin{deluxetable*}{cccccc}
\tabletypesize{\scriptsize}
\tablecaption{The integrated stellar masses and SFRs of HAEs and DRGs.}
\tablewidth{0pt}
\tablehead{
\colhead{Region} & \multicolumn{3}{c}{Stellar mass ($\Sigma$ \Mstar) [\Msun]} & \colhead{SFR ($\Sigma$ {\rm SFR}) [\Msun\ yr$^{-1}$]} &
\colhead{Specific SFR ($\Sigma$ {\rm SFR} / $\Sigma$ \Mstar) [yr$^{-1}$]} \\
\cline{2-4}\\[-1mm]
\colhead{}       & \colhead{HAE(blue)} & \colhead{HAE(red)} &
\colhead{DRG} & \colhead{HAE} & \colhead{HAE+DRG}
}
\startdata
Clump 1      & 1.3$\times10^{11}$ & 2.3$\times10^{11}$ & 1.2$\times10^{12}$ & 1.8$\times10^{3}$ & 1.4$\times10^{-9}$ \\
Clump 2      & 3.9$\times10^{11}$ & 2.7$\times10^{11}$ & 1.2$\times10^{12}$ & 3.4$\times10^{3}$ & 2.2$\times10^{-9}$ \\
Clump 3      & 8.2$\times10^{10}$ & 9.1$\times10^{10}$ & 1.1$\times10^{11}$ & 1.0$\times10^{3}$ & 5.4$\times10^{-9}$ \\
All clumps   & 6.1$\times10^{11}$ & 6.0$\times10^{11}$ & 2.2$\times10^{12}$ & 6.2$\times10^{3}$ & 2.2$\times10^{-9}$ \\
Others       & 4.1$\times10^{11}$ & 1.8$\times10^{11}$ & 9.2$\times10^{11}$ & 3.2$\times10^{3}$ & 2.4$\times10^{-9}$ \\
Entire field & 1.0$\times10^{12}$ & 7.8$\times10^{11}$ & 2.8$\times10^{12}$ & 9.4$\times10^{3}$ & 2.4$\times10^{-9}$
\enddata
\tablecomments{The numbers are also divided into clumps and to the other region.
  For HAEs, blue and red ones are separated at $J-K_s=1.38$ (same as the DRG criterion).
  For DRGs, the field contamination from foreground/background fields along the
  line of sight is statistically subtracted using the general field data (MODS).
  In estimation of specific SFR for combined samples with HAE and DRG,
  stellar masses are calculated by summing up those for HAE(blue) and DRG.
  The radio galaxy in the clump-1 is excluded in this table,
  because its \ha\ flux is severely contaminated by the AGN component.
  }
\label{table;masssfr}
\end{deluxetable*}

\section{Discussions}
\label{sec;discussions}

\subsection{red \ha\ emitters}
\label{sec;redHAE}
\begin{figure}
\epsscale{1.3}
\plotone{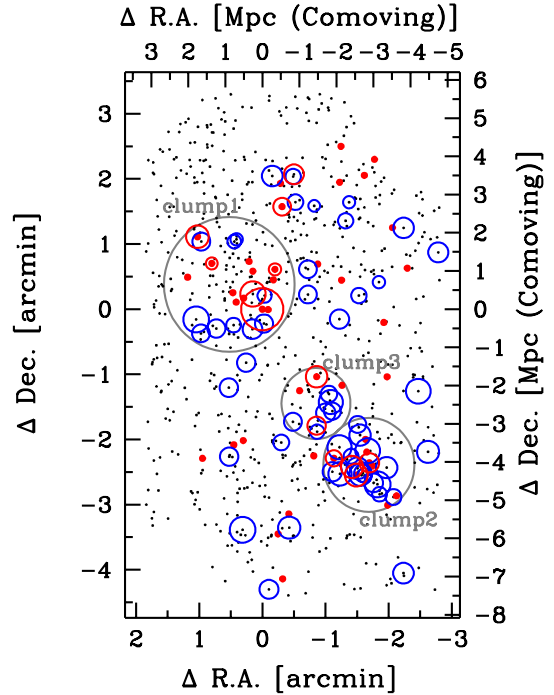}
\caption{Similar to Figure \ref{fig;map_HAE}, but the size of the circles
  is now scaled with SFR, in the sense that larger
  symbols show higher SFRs of the \ha\ emitters. The red \ha\ emitters
  with $J-Ks>1.38$ (DRGs criterion) are also specified with red circles.
  The red filled dots show DRGs. 
  \label{fig;map_redHAE}}
\end{figure}

As described in \S~\ref{sec;CMD}, there are some HAEs with colors
redder than $J-K_s$=1.38. Figure~\ref{fig;map_redHAE} shows the
distribution of the HAEs and DRGs, where the red HAEs with
$J-K_s>1.38$ are marked by red open circles. It seems that the red
HAEs are clustered around the radio galaxy (clump-1) and in the
south-west clumps (clump-2 and clump-3). Such concentration of the red
emitters in high density regions is qualitatively different from what
is seen in the lower redshift cluster at $z\sim0.8$ where red HAEs
prefer to medium density regions, as mentioned in
\S~\ref{sec;CMD}. This may indicate that, in this proto-cluster at
$z\sim2.5$, the transition of galaxies is occurring in the densest
environment rather than in the outskirts. We note that most of the red
HAEs are less massive than $10^{11}$\Msun\ as shown in
Figure~\ref{fig;CMD}. If the dusty starburst phase does not last so
long, these red HAEs would stay on relatively faint end of the red
sequence even after they quench star formation. If the DRGs without
\ha\ emission lines with stellar masses of a few $\times 10^{11}$\Msun
are really proto-cluster members, the active phase of these massive
quiescent galaxies should have been seen at $z\gtrsim3$. Some of their
progenitors would be like the populations of LBGs or LAEs which are in
fact often identified in over-dense regions (proto-clusters) at
$z\gtrsim3$ \citep[e.g.,][]{steidel1998,Miley2004,Venemans2005,Venemans2007,Ouchi2005,Kuiper2011}.  

On the other hand, as another possibility, the red HAEs can be
quiescent galaxies with AGN activity in the galaxy core, not dusty
starburst galaxies. Even in that case, the fact that the red HAEs are
clustered in high density region is very interesting. This is because
the considered situation suggests that AGN activity is enhanced in the
core regions of the clumps, and thus AGN must give a significant
influence on the evolution of high-$z$ galaxies in high density
region. Perhaps, AGN activity has something to do with process to
quench the star formation of cluster galaxies, i.e., so-called AGN
feedback. Although it is difficult to reveal which hypotheses are true
between dust starbursts and quiescent galaxies hosting an AGN with
only the data now available, in any case, there is no doubt that
galaxies are activated in the densest regions such as clumps at 
$z\sim2.5$.  

In lower-$z$ clusters ($z\lesssim1.0$), the fraction of star forming
galaxies to all cluster members decreases as we go closer to cluster
centers or as the number density of galaxies increases
\citep[e.g.,][]{kodama2004,koyama2010,koyama2011}.  
In this USS~1558-003 proto-cluster, however, it is difficult to
completely sample red quiescent galaxies, since DRGs have broad
redshift distribution such as $2 \lesssim z \lesssim 4$ and also
contain both populations of quiescent galaxies and dusty starbursts.
This makes it hard for us to quantify the fraction of star forming
galaxies in this proto-cluster. However, even if we assume that all
the DRGs are quiescent galaxies associated to the proto-cluster at
$z=2.53$, a large fraction of proto-cluster members are star-formers,
and there are much fewer red quiescent galaxies.

\subsection{extended \ha\ emission in radio galaxy}
\label{sec;radiogalaxy}
As shown in Figure~\ref{fig;haRG}, the radio galaxy USS~1558-003 has
quite an extended \ha\ emission with the apparent scale of
$\sim4.5$\arcsec\ or the physical scale of ~$\sim$36 kpc. Although the
spatially extended nebular line emission (\lya\ and \ha) have already
been recognized around this radio galaxy
\citep{VillarMartn2007,Humphrey2008}, we here reveal for the first
time the two-dimensional distribution of \ha\ emission.
Such extended nebular emission is a remarkable feature commonly seen
in HzRGs \citep[e.g.,][]{Heckman1991,Humphrey2007}. 
In particular, it is well-known that they show extended \lya\ emission
aligned with radio structure (alignment effect). Our extended \ha\
emission is tilted with the position angle (PA) of 50\arcdeg\
(Figure~\ref{fig;haRG}). This direction is approximately aligned to
but slightly different from the radio axis which has the PA of
75\arcdeg\ \citep{Pentericci2000}. Based on the integral field
spectroscopy, \citet{VillarMartn2007} found that the extent of the
\lya\ emission is $\sim$9.0\arcsec$\times$7.5\arcsec, and is
misaligned by $\sim$30$\pm$5\arcdeg\ relatively to the radio
structures. Therefore, the direction of the stretch of \ha\ emission
is more or less similar to that of \lya\ emission. However, the size
of \ha\ emission is $\sim4.5\arcsec\times3.0\arcsec$, and is much
smaller than the \lya\ emission. 
It is possible that this smaller size of \ha\ emission is due to the
fact that surface brightness of the outer region is less than the
detection limit. To check the possibility, we compare the distribution
of \lya\ flux density given in \citet{VillarMartn2007} with the
NB2315--$K_s$ image and find that the \ha\ emission can be detected at
more than 2$\sigma$ if \lya/\ha\ ratio is constant over the whole
region where \lya\ emission is detected. This thus means the ratio of
\lya/\ha\ is not constant but larger in the outer region of the radio
galaxy. Such smaller size of \ha\ emission compared to \lya\ emission
is also seen in the radio galaxy MRC~1138-262 at $z$=2.16
\citep{kurk2002}. \citet{kurk2002} conclude that such fairly extended
\lya\ emission is due to scattered radiation. Since we see a similar
nature in the USS~1558-003 radio galaxy, the existence of abundant
scattering material around the radio galaxy trapped in deep potential
wells and large \lya\ emission probably due to resonant scattering
seem to be a common characteristic of HzRGs.  

Furthermore, we find that the radio galaxy also has extended emission
in $B$ and $r'$-band, where there seems to be the contribution of
\lya\ and He {\scriptsize II} emissions to some extend in the $B$ and
$r'$-band, respectively. However, we do not see any extended emission
in the $K_s$-band. Such extended emission in the rest-frame UV
wavelengths, and the compactness of old stellar component may suggest
that star forming activity is also currently on-going in the extended
regions. In that case, such extended star formation is likely
triggered by some physical processes associated to the radio
activity. The jet-induced star formation is one of the plausible
options \citep[e.g.,][]{Bicknell2000,hatch2011}.       

Many studies have been conducted to investigate the physical origin of
nebular emission around the radio galaxy. Possible mechanisms include
the photoionization by AGN, young stars and X-ray emission from shock
heated gas, and collisional ionization by jet-induced shock
\citep[e.g.,][and references therein]{Miley2008}.     
Although the ionization mechanism has not yet been completely
understood, recent studies suggest that AGN could be the dominant
source to ionize the surrounding nebular gas
\citep[e.g.,][]{Humphrey2008,Miley2008}. However, for radio galaxies
at $z\sim1$, \citet{Best2000} found that the cause of nebular emission
is dependent on the size of radio morphology based on the diagnostic
with line ratios of nebular emission. Small sources with a radio
structure smaller than $\lesssim150$ kpc tend to have emission
originated from shock ionization, while larger sources prefer the
mechanism of photoionization by AGNs. Since the radio observation of
USS~1558-003 shows that the size of the radio galaxy is 9.2\arcsec\
(74kpc) \citep{Pentericci2000}, the shock ionization would be
preferred. It is therefore interesting to investigate the origin of
\ha\ emission of this radio galaxy more in detail.

For this purpose, a diagnostic with nebular emission in rest-frame
optical wavelength is useful. \citet{Humphrey2008} conducted optical
and NIR spectroscopies where a slit was placed on the radio galaxy
along the radio axis, and investigated the ratios of detected emission
lines. They detected a lot of emission lines in the rest-frame
wavelength range of $\lambda=$1216--6585\AA, and found that the radio
galaxy has a broad \ha\ emission line with a velocity width of 12000
km s$^{-1}$, and that the amount of dust is negligibly small (A$_{\rm
  v}\sim0$). Consequently, they concluded that the line ratios were
best explained by AGN photoionization. However, they also find that
the sources with high A$_{\rm v}$ do not show evidence for jet-gas
interactions, while the sources undergoing strong jet-gas interactions
have low A$_{\rm v}$. The small amount of dust in this radio galaxy
implies that the surrounding gas is undergoing interaction with the
jet-induced shock. If this is indeed the case, the result is
consistent with the proposed relation between the ionization mechanism
and the radio size. It is suggestive therefore that the shock induced
star formation is occurring to some extent as well as AGN
photoionization, and thus both mechanisms are probably contributing to
the large extent of the \ha\ emission. It is difficult, however, to
quantify the relative contribution of the two mechanisms, and it is
beyond the scope of this paper. Integral field spectroscopy at NIR is
essential to resolve the two-dimensional structure of the emission
line regions and to fully understand the physical origins of nebular
emission. 

\section{Conclusions}
\label{sec;conclusions}
We have conducted a panoramic narrow-band imaging of \ha\ emitters (HAEs)
in the proto-cluster candidate around the radio galaxy USS~1558-003
at $z=2.53$ using NB2315 filter ($\lambda_c=2.313\micron, \Delta\lambda=0.027\micron$)
installed in MOIRCS on Subaru Telescope.
This target is known as an over-dense region where
distant red galaxies (DRGs) are clustered \citep{kodama2007}.
We have confirmed that this is indeed a rich proto-cluster in making
with lots of star forming galaxies (HAEs) associated
to the radio galaxy.
We have mapped out the 2-D structure of the proto-cluster, and investigated
the star forming activities and the stellar mass content of this forming
cluster. The main results we have found are summarized below.

\begin{enumerate}[(i)]

\item The proto-cluster is mainly composed of three conspicuous groups of galaxies.
    One of them is surrounding the radio galaxy, and another is about 1.5 Mpc
    (physical scale) away from the radio galaxy to the south-west, and the other is in between
    the two clumps. These groups show significant excess in the number densities of
    both HAEs and DRGs.
    Their close separations suggest that they would merge together in the near future
    and grow to a single, more massive galaxy cluster at later times. 

\item A large fraction of the \ha\ emitters in this proto-cluster have SFRs
    higher than 100 \Msun\ yr$^{-1}$, indicating that at $z\sim2.5$, the progenitors
    of cluster early-type galaxies are vigorously forming in the biased high
    density regions.
    Star formation activity is high everywhere irrespective of environment
    within the proto-cluster region, and the properties of individual HAEs
    show little environmental dependence, except that the HAEs in the densest
    clump may have slightly higher star formation rates compared to those in
    other regions.

\item Most of the \ha\ emitters have blue colors, but some emitters have
    very red colors comparable to DRG (i.e., $J-Ks>1.38$).
    Those red emitters are located on the fainter side of the
    red sequence on the color-magnitude diagram, except for the radio galaxy itself.
    Moreover, the red \ha\ emitters tend to be clustered in the three highest
    density clumps in contrast to lower-$z$ clusters where
    similar red emitters are avoiding the cluster cores and preferentially
    located in the medium density regions or the outskirts of the clusters.
    Since the red emitters are likely to be dusty starburst galaxies
    in the transitional phase, this result may indicate that some
    environmental effects, such as galaxy-galaxy interaction, are at
    work on galaxies in the dense proto-cluster core at $z=2.53$ and
    they are just changing their properties rapidly. 

\item The radio galaxy shows a large \ha\ halo extended over $\sim$4.5\arcsec\ (i.e.,
    36 kpc), indicating that some ionization mechanisms (AGN, young stars, and shocks)
    are at work in the surrounding material around the radio galaxy.
    Such spatial extent of \ha\ emission is still much smaller than the \lya\ halo.
    This means that the \lya\ emission is severely extended by resonant scattering.

\end{enumerate}

These results we present in this paper are all intriguing and 
there is no doubt that proto-clusters at $z>2$ have high star
forming activity and are in vigorously evolving phase.
However, we need to investigate more samples of proto-clusters
at $z>2$ in detail in order to answer the question whether our
results represent the universal properties of proto-clusters at
$z>2$ or showing the specific characteristics of the USS1558 proto-cluster.
Our on-going MAHALO-Subaru project has been surveying other
proto-clusters as well as un-biasedly selected fields at similar
redshifts, and enabling us to reveal the universality or variation
of properties among proto-clusters at $z>2$. Such discussion will
be presented in our future papers (e.g., Koyama et al. in prep.).

\acknowledgments
We would like to thank an anonymous referee for carefully reading our
manuscript and giving useful comments. All of the data used in this
paper are collected at Subaru Telescope, which is operated by the
National Astronomical Observatory of Japan. We thank the Subaru
Telescope staff for their invaluable help to assist our observations
with Suprime-Cam and MOIRCS. We acknowledge Dr.\ Philip Best for
useful discussion. MH is grateful for the financial support from the
Japan Society for the Promotion of Science (JSPS) fund,
``Institutional Program for Young Researcher Overseas Visits'' to stay
IfA, Royal Observatory of Edinburgh for two months.

{\it Facilities:} \facility{Subaru}.



\begin{thebibliography}{85}
\expandafter\ifx\csname natexlab\endcsname\relax\def\natexlab#1{#1}\fi

\bibitem[{{Bertin} \& {Arnouts}(1996)}]{bertin1996}
{Bertin}, E., \& {Arnouts}, S. 1996, A\&AS, 117, 393

\bibitem[{{Best} {et~al.}(2000){Best}, {R{\"o}ttgering}, \&
  {Longair}}]{Best2000}
{Best}, P.~N., {R{\"o}ttgering}, H.~J.~A., \& {Longair}, M.~S. 2000, \mnras,
  311, 23

\bibitem[{{Bicknell} {et~al.}(2000){Bicknell}, {Sutherland}, {van Breugel},
  {Dopita}, {Dey}, \& {Miley}}]{Bicknell2000}
{Bicknell}, G.~V., {Sutherland}, R.~S., {van Breugel}, W.~J.~M., {Dopita},
  M.~A., {Dey}, A., \& {Miley}, G.~K. 2000, \apj, 540, 678

\bibitem[{{Brammer} {et~al.}(2008){Brammer}, {van Dokkum}, \&
  {Coppi}}]{brammer2008}
{Brammer}, G.~B., {van Dokkum}, P.~G., \& {Coppi}, P. 2008, ApJ, 686, 1503

\bibitem[{{Bruzual} \& {Charlot}(2003)}]{bc03}
{Bruzual}, G., \& {Charlot}, S. 2003, \mnras, 344, 1000

\bibitem[{{Butcher} \& {Oemler}(1978)}]{butcher1978}
{Butcher}, H., \& {Oemler}, Jr., A. 1978, \apj, 219, 18

\bibitem[{{Butcher} \& {Oemler}(1984)}]{butcher1984}
---. 1984, \apj, 285, 426

\bibitem[{{Byrd} \& {Valtonen}(1990)}]{Byrd1990}
{Byrd}, G., \& {Valtonen}, M. 1990, \apj, 350, 89

\bibitem[{{Calzetti} {et~al.}(2000){Calzetti}, {Armus}, {Bohlin}, {Kinney},
  {Koornneef}, \& {Storchi-Bergmann}}]{calzetti2000}
{Calzetti}, D., {Armus}, L., {Bohlin}, R.~C., {Kinney}, A.~L., {Koornneef}, J.,
  \& {Storchi-Bergmann}, T. 2000, ApJ, 533, 682

\bibitem[{{Cardelli} {et~al.}(1989){Cardelli}, {Clayton}, \&
  {Mathis}}]{cardelli1989}
{Cardelli}, J.~A., {Clayton}, G.~C., \& {Mathis}, J.~S. 1989, ApJ, 345, 245

\bibitem[{{Coleman} {et~al.}(1980){Coleman}, {Wu}, \& {Weedman}}]{coleman1980}
{Coleman}, G.~D., {Wu}, C., \& {Weedman}, D.~W. 1980, ApJS, 43, 393

\bibitem[{{Cooper} {et~al.}(2008){Cooper}, {Newman}, {Weiner}, {Yan},
  {Willmer}, {Bundy}, {Coil}, {Conselice}, {Davis}, {Faber}, {Gerke},
  {Guhathakurta}, {Koo}, \& {Noeske}}]{cooper2008}
{Cooper}, M.~C., {et~al.} 2008, \mnras, 383, 1058

\bibitem[{{Cucciati} {et~al.}(2006){Cucciati}, {Iovino}, {Marinoni}, {Ilbert},
  {Bardelli}, {Franzetti}, {Le F{\`e}vre}, {Pollo}, {Zamorani}, {Cappi},
  {Guzzo}, {McCracken}, {Meneux}, {Scaramella}, {Scodeggio}, {Tresse}, {Zucca},
  {Bottini}, {Garilli}, {Le Brun}, {Maccagni}, {Picat}, {Vettolani},
  {Zanichelli}, {Adami}, {Arnaboldi}, {Arnouts}, {Bolzonella}, {Charlot},
  {Ciliegi}, {Contini}, {Foucaud}, {Gavignaud}, {Marano}, {Mazure}, {Merighi},
  {Paltani}, {Pell{\`o}}, {Pozzetti}, {Radovich}, {Bondi}, {Bongiorno},
  {Busarello}, {de la Torre}, {Gregorini}, {Lamareille}, {Mathez}, {Mellier},
  {Merluzzi}, {Ripepi}, {Rizzo}, {Temporin}, \& {Vergani}}]{cucciati2006}
{Cucciati}, O., {et~al.} 2006, \aap, 458, 39

\bibitem[{{Daddi} {et~al.}(2004){Daddi}, {Cimatti}, {Renzini}, {Fontana},
  {Mignoli}, {Pozzetti}, {Tozzi}, \& {Zamorani}}]{daddi2004}
{Daddi}, E., {Cimatti}, A., {Renzini}, A., {Fontana}, A., {Mignoli}, M.,
  {Pozzetti}, L., {Tozzi}, P., \& {Zamorani}, G. 2004, ApJ, 617, 746

\bibitem[{{Daddi} {et~al.}(2007){Daddi}, {Dickinson}, {Morrison}, {Chary},
  {Cimatti}, {Elbaz}, {Frayer}, {Renzini}, {Pope}, {Alexander}, {Bauer},
  {Giavalisco}, {Huynh}, {Kurk}, \& {Mignoli}}]{Daddi2007}
{Daddi}, E., {et~al.} 2007, ApJ, 670, 156

\bibitem[{{Doherty} {et~al.}(2010){Doherty}, {Tanaka}, {De Breuck}, {Ly},
  {Kodama}, {Kurk}, {Seymour}, {Vernet}, {Stern}, {Venemans}, {Kajisawa}, \&
  {Tanaka}}]{doherty2010}
{Doherty}, M., {et~al.} 2010, A\&A, 509, A83+

\bibitem[{{Dressler} {et~al.}(1997){Dressler}, {Oemler}, {Couch}, {Smail},
  {Ellis}, {Barger}, {Butcher}, {Poggianti}, \& {Sharples}}]{dressler1997}
{Dressler}, A., {et~al.} 1997, \apj, 490, 577

\bibitem[{{Fassbender} {et~al.}(2011){Fassbender}, {Nastasi}, {B{\"o}hringer},
  {{\v S}uhada}, {Santos}, {Rosati}, {Pierini}, {M{\"u}hlegger}, {Quintana},
  {Schwope}, {Lamer}, {de Hoon}, {Kohnert}, {Pratt}, \&
  {Mohr}}]{fassbender2011}
{Fassbender}, R., {et~al.} 2011, A\&A, 527, L10+

\bibitem[{{Franx} {et~al.}(2003){Franx}, {Labb{\'e}}, {Rudnick}, {van Dokkum},
  {Daddi}, {F{\"o}rster Schreiber}, {Moorwood}, {Rix}, {R{\"o}ttgering}, {van
  der Wel}, {van der Werf}, \& {van Starkenburg}}]{Franx2003}
{Franx}, M., {et~al.} 2003, \apjl, 587, L79

\bibitem[{{Garn} {et~al.}(2010){Garn}, {Sobral}, {Best}, {Geach}, {Smail},
  {Cirasuolo}, {Dalton}, {Dunlop}, {McLure}, \& {Farrah}}]{Garn2010}
{Garn}, T., {et~al.} 2010, MNRAS, 402, 2017

\bibitem[{{Geach} {et~al.}(2008){Geach}, {Smail}, {Best}, {Kurk}, {Casali},
  {Ivison}, \& {Coppin}}]{geach2008}
{Geach}, J.~E., {Smail}, I., {Best}, P.~N., {Kurk}, J., {Casali}, M., {Ivison},
  R.~J., \& {Coppin}, K. 2008, \mnras, 388, 1473

\bibitem[{{Geach} {et~al.}(2006){Geach}, {Smail}, {Ellis}, {Moran}, {Smith},
  {Treu}, {Kneib}, {Edge}, \& {Kodama}}]{geach2006}
{Geach}, J.~E., {et~al.} 2006, \apj, 649, 661

\bibitem[{{Giavalisco} {et~al.}(2004){Giavalisco}, {Ferguson}, {Koekemoer},
  {Dickinson}, {Alexander}, {Bauer}, {Bergeron}, {Biagetti}, {Brandt},
  {Casertano}, {Cesarsky}, {Chatzichristou}, {Conselice}, {Cristiani}, {Da
  Costa}, {Dahlen}, {de Mello}, {Eisenhardt}, {Erben}, {Fall}, {Fassnacht},
  {Fosbury}, {Fruchter}, {Gardner}, {Grogin}, {Hook}, {Hornschemeier}, {Idzi},
  {Jogee}, {Kretchmer}, {Laidler}, {Lee}, {Livio}, {Lucas}, {Madau},
  {Mobasher}, {Moustakas}, {Nonino}, {Padovani}, {Papovich}, {Park},
  {Ravindranath}, {Renzini}, {Richardson}, {Riess}, {Rosati}, {Schirmer},
  {Schreier}, {Somerville}, {Spinrad}, {Stern}, {Stiavelli}, {Strolger},
  {Urry}, {Vandame}, {Williams}, \& {Wolf}}]{Giavalisco2004}
{Giavalisco}, M., {et~al.} 2004, \apjl, 600, L93

\bibitem[{{Gobat} {et~al.}(2011){Gobat}, {Daddi}, {Onodera}, {Finoguenov},
  {Renzini}, {Arimoto}, {Bouwens}, {Brusa}, {Chary}, {Cimatti}, {Dickinson},
  {Kong}, \& {Mignoli}}]{gobat2011}
{Gobat}, R., {et~al.} 2011, A\&A, 526, A133+

\bibitem[{{Gunn} \& {Gott}(1972)}]{Gunn1972}
{Gunn}, J.~E., \& {Gott}, III, J.~R. 1972, \apj, 176, 1

\bibitem[{{Gunn} \& {Stryker}(1983)}]{gunn1983}
{Gunn}, J.~E., \& {Stryker}, L.~L. 1983, ApJS, 52, 121

\bibitem[{{Hatch} {et~al.}(2011){Hatch}, {Kurk}, {Pentericci}, {Venemans},
  {Kuiper}, {Miley}, \& {R{\"o}ttgering}}]{hatch2011}
{Hatch}, N.~A., {Kurk}, J.~D., {Pentericci}, L., {Venemans}, B.~P., {Kuiper},
  E., {Miley}, G.~K., \& {R{\"o}ttgering}, H.~J.~A. 2011, \mnras, 1103

\bibitem[{{Hatch} {et~al.}(2009){Hatch}, {Overzier}, {Kurk}, {Miley},
  {R{\"o}ttgering}, \& {Zirm}}]{Hatch2009}
{Hatch}, N.~A., {Overzier}, R.~A., {Kurk}, J.~D., {Miley}, G.~K.,
  {R{\"o}ttgering}, H.~J.~A., \& {Zirm}, A.~W. 2009, \mnras, 395, 114

\bibitem[{{Hayashi} {et~al.}(2011){Hayashi}, {Kodama}, {Koyama}, {Tadaki}, \&
  {Tanaka}}]{hayashi2011}
{Hayashi}, M., {Kodama}, T., {Koyama}, Y., {Tadaki}, K.-i., \& {Tanaka}, I.
  2011, ArXiv e-prints

\bibitem[{{Hayashi} {et~al.}(2010){Hayashi}, {Kodama}, {Koyama}, {Tanaka},
  {Shimasaku}, \& {Okamura}}]{hayashi2010}
{Hayashi}, M., {Kodama}, T., {Koyama}, Y., {Tanaka}, I., {Shimasaku}, K., \&
  {Okamura}, S. 2010, \mnras, 402, 1980

\bibitem[{{Heckman} {et~al.}(1991){Heckman}, {Miley}, {Lehnert}, \& {van
  Breugel}}]{Heckman1991}
{Heckman}, T.~M., {Miley}, G.~K., {Lehnert}, M.~D., \& {van Breugel}, W. 1991,
  \apj, 370, 78

\bibitem[{{Hilton} {et~al.}(2010){Hilton}, {Lloyd-Davies}, {Stanford}, {Stott},
  {Collins}, {Romer}, {Hosmer}, {Hoyle}, {Kay}, {Liddle}, {Mehrtens}, {Miller},
  {Sahl{\'e}n}, \& {Viana}}]{hilton2010}
{Hilton}, M., {et~al.} 2010, ApJ, 718, 133

\bibitem[{{Hopkins} \& {Beacom}(2006)}]{hopkins2006}
{Hopkins}, A.~M., \& {Beacom}, J.~F. 2006, ApJ, 651, 142

\bibitem[{{Humphrey} {et~al.}(2007){Humphrey}, {Villar-Mart{\'{\i}}n},
  {Fosbury}, {Binette}, {Vernet}, {De Breuck}, \& {di Serego
  Alighieri}}]{Humphrey2007}
{Humphrey}, A., {Villar-Mart{\'{\i}}n}, M., {Fosbury}, R., {Binette}, L.,
  {Vernet}, J., {De Breuck}, C., \& {di Serego Alighieri}, S. 2007, \mnras,
  375, 705

\bibitem[{{Humphrey} {et~al.}(2008){Humphrey}, {Villar-Mart{\'{\i}}n},
  {Vernet}, {Fosbury}, {di Serego Alighieri}, \& {Binette}}]{Humphrey2008}
{Humphrey}, A., {Villar-Mart{\'{\i}}n}, M., {Vernet}, J., {Fosbury}, R., {di
  Serego Alighieri}, S., \& {Binette}, L. 2008, \mnras, 383, 11

\bibitem[{{Ichikawa} {et~al.}(2006){Ichikawa}, {Suzuki}, {Tokoku}, {Uchimoto},
  {Konishi}, {Yoshikawa}, {Yamada}, {Tanaka}, {Omata}, \&
  {Nishimura}}]{ichikawa2006}
{Ichikawa}, T., {et~al.} 2006, in Society of Photo-Optical Instrumentation
  Engineers (SPIE) Conference Series, Vol. 6269, Society of Photo-Optical
  Instrumentation Engineers (SPIE) Conference Series

\bibitem[{{Kajisawa} {et~al.}(2006{\natexlab{a}}){Kajisawa}, {Kodama},
  {Tanaka}, {Yamada}, \& {Bower}}]{kajisawa2006}
{Kajisawa}, M., {Kodama}, T., {Tanaka}, I., {Yamada}, T., \& {Bower}, R.
  2006{\natexlab{a}}, MNRAS, 371, 577

\bibitem[{{Kajisawa} {et~al.}(2006{\natexlab{b}}){Kajisawa}, {Konishi},
  {Suzuki}, {Tokoku}, {Uchimoto}, {Katsuno}, {Yoshikawa}, {Akiyama},
  {Ichikawa}, {Ouchi}, {Omata}, {Tanaka}, {Nishimura}, \&
  {Yamada}}]{Kajisawa2006PASJ}
{Kajisawa}, M., {et~al.} 2006{\natexlab{b}}, \pasj, 58, 951

\bibitem[{{Kajisawa} {et~al.}(2011){Kajisawa}, {Ichikawa}, {Tanaka}, {Yamada},
  {Akiyama}, {Suzuki}, {Tokoku}, {Katsuno Uchimoto}, {Konishi}, {Yoshikawa},
  {Nishimura}, {Omata}, {Ouchi}, {Iwata}, {Hamana}, \&
  {Onodera}}]{Kajisawa2011}
---. 2011, \pasj, 63, 379

\bibitem[{{Kauffmann} {et~al.}(2004){Kauffmann}, {White}, {Heckman},
  {M{\'e}nard}, {Brinchmann}, {Charlot}, {Tremonti}, \&
  {Brinkmann}}]{kauffmann2004}
{Kauffmann}, G., {White}, S.~D.~M., {Heckman}, T.~M., {M{\'e}nard}, B.,
  {Brinchmann}, J., {Charlot}, S., {Tremonti}, C., \& {Brinkmann}, J. 2004,
  \mnras, 353, 713

\bibitem[{{Kennicutt}(1998)}]{kennicutt1998}
{Kennicutt}, Jr., R.~C. 1998, ARA\&A, 36, 189

\bibitem[{{Kodama} {et~al.}(1998){Kodama}, {Arimoto}, {Barger}, \&
  {Arag'on-Salamanca}}]{kodama1998}
{Kodama}, T., {Arimoto}, N., {Barger}, A.~J., \& {Arag'on-Salamanca}, A. 1998,
  A\&A, 334, 99

\bibitem[{{Kodama} {et~al.}(2004){Kodama}, {Balogh}, {Smail}, {Bower}, \&
  {Nakata}}]{kodama2004}
{Kodama}, T., {Balogh}, M.~L., {Smail}, I., {Bower}, R.~G., \& {Nakata}, F.
  2004, MNRAS, 354, 1103

\bibitem[{{Kodama} {et~al.}(1999){Kodama}, {Bell}, \& {Bower}}]{Kodama1999}
{Kodama}, T., {Bell}, E.~F., \& {Bower}, R.~G. 1999, \mnras, 302, 152

\bibitem[{{Kodama} {et~al.}(2007){Kodama}, {Tanaka}, {Kajisawa}, {Kurk},
  {Venemans}, {De Breuck}, {Vernet}, \& {Lidman}}]{kodama2007}
{Kodama}, T., {Tanaka}, I., {Kajisawa}, M., {Kurk}, J., {Venemans}, B., {De
  Breuck}, C., {Vernet}, J., \& {Lidman}, C. 2007, MNRAS, 377, 1717

\bibitem[{{Kong} {et~al.}(2006){Kong}, {Daddi}, {Arimoto}, {Renzini},
  {Broadhurst}, {Cimatti}, {Ikuta}, {Ohta}, {da Costa}, {Olsen}, {Onodera}, \&
  {Tamura}}]{kong2006}
{Kong}, X., {et~al.} 2006, \apj, 638, 72

\bibitem[{{Koyama} {et~al.}(2011){Koyama}, {Kodama}, {Nakata}, {Shimasaku}, \&
  {Okamura}}]{koyama2011}
{Koyama}, Y., {Kodama}, T., {Nakata}, F., {Shimasaku}, K., \& {Okamura}, S.
  2011, \apj, 734, 66

\bibitem[{{Koyama} {et~al.}(2010){Koyama}, {Kodama}, {Shimasaku}, {Hayashi},
  {Okamura}, {Tanaka}, \& {Tokoku}}]{koyama2010}
{Koyama}, Y., {Kodama}, T., {Shimasaku}, K., {Hayashi}, M., {Okamura}, S.,
  {Tanaka}, I., \& {Tokoku}, C. 2010, MNRAS, 403, 1611

\bibitem[{{Koyama} {et~al.}(2008){Koyama}, {Kodama}, {Shimasaku}, {Okamura},
  {Tanaka}, {Lee}, {Im}, {Matsuhara}, {Takagi}, {Wada}, \&
  {Oyabu}}]{koyama2008}
{Koyama}, Y., {et~al.} 2008, MNRAS, 391, 1758

\bibitem[{{Kriek} {et~al.}(2008){Kriek}, {van der Wel}, {van Dokkum}, {Franx},
  \& {Illingworth}}]{kriek2008}
{Kriek}, M., {van der Wel}, A., {van Dokkum}, P.~G., {Franx}, M., \&
  {Illingworth}, G.~D. 2008, ApJ, 682, 896

\bibitem[{{Kron}(1980)}]{Kron1980}
{Kron}, R.~G. 1980, \apjs, 43, 305

\bibitem[{{Kuiper} {et~al.}(2010){Kuiper}, {Hatch}, {R{\"o}ttgering}, {Miley},
  {Overzier}, {Venemans}, {De Breuck}, {Croft}, {Kajisawa}, {Kodama}, {Kurk},
  {Pentericci}, {Stanford}, {Tanaka}, \& {Zirm}}]{Kuiper2010}
{Kuiper}, E., {et~al.} 2010, \mnras, 405, 969

\bibitem[{{Kuiper} {et~al.}(2011){Kuiper}, {Hatch}, {Venemans}, {Miley},
  {R{\"o}ttgering}, {Kurk}, {Overzier}, {Pentericci}, {Bland-Hawthorn}, \&
  {Cepa}}]{Kuiper2011}
---. 2011, \mnras, 417, 1088

\bibitem[{{Kurk} {et~al.}(2004{\natexlab{a}}){Kurk}, {Pentericci}, {Overzier},
  {R{\"o}ttgering}, \& {Miley}}]{kurk2004b}
{Kurk}, J.~D., {Pentericci}, L., {Overzier}, R.~A., {R{\"o}ttgering}, H.~J.~A.,
  \& {Miley}, G.~K. 2004{\natexlab{a}}, \aap, 428, 817

\bibitem[{{Kurk} {et~al.}(2002){Kurk}, {Pentericci}, {R{\"o}ttgering}, \&
  {Miley}}]{kurk2002}
{Kurk}, J.~D., {Pentericci}, L., {R{\"o}ttgering}, H.~J.~A., \& {Miley}, G.~K.
  2002, in Revista Mexicana de Astronomia y Astrofisica Conference Series,
  Vol.~13, Revista Mexicana de Astronomia y Astrofisica Conference Series, ed.
  {W.~J.~Henney, W.~Steffen, L.~Binette, \& A.~Raga}, 191--195

\bibitem[{{Kurk} {et~al.}(2004{\natexlab{b}}){Kurk}, {Pentericci},
  {R{\"o}ttgering}, \& {Miley}}]{kurk2004a}
{Kurk}, J.~D., {Pentericci}, L., {R{\"o}ttgering}, H.~J.~A., \& {Miley}, G.~K.
  2004{\natexlab{b}}, \aap, 428, 793

\bibitem[{{Ly} {et~al.}(2007){Ly}, {Malkan}, {Kashikawa}, {Shimasaku}, {Doi},
  {Nagao}, {Iye}, {Kodama}, {Morokuma}, \& {Motohara}}]{ly2007}
{Ly}, C., {et~al.} 2007, ApJ, 657, 738

\bibitem[{{Madau}(1995)}]{madau1995}
{Madau}, P. 1995, \apj, 441, 18

\bibitem[{{Madau} {et~al.}(1996){Madau}, {Ferguson}, {Dickinson}, {Giavalisco},
  {Steidel}, \& {Fruchter}}]{madau1996}
{Madau}, P., {Ferguson}, H.~C., {Dickinson}, M.~E., {Giavalisco}, M.,
  {Steidel}, C.~C., \& {Fruchter}, A. 1996, \mnras, 283, 1388

\bibitem[{{McLure} {et~al.}(1999){McLure}, {Kukula}, {Dunlop}, {Baum}, {O'Dea},
  \& {Hughes}}]{McLure1999}
{McLure}, R.~J., {Kukula}, M.~J., {Dunlop}, J.~S., {Baum}, S.~A., {O'Dea},
  C.~P., \& {Hughes}, D.~H. 1999, \mnras, 308, 377

\bibitem[{{Miley} \& {De Breuck}(2008)}]{Miley2008}
{Miley}, G., \& {De Breuck}, C. 2008, \aapr, 15, 67

\bibitem[{{Miley} {et~al.}(2004){Miley}, {Overzier}, {Tsvetanov}, {Bouwens},
  {Ben{\'{\i}}tez}, {Blakeslee}, {Ford}, {Illingworth}, {Postman}, {Rosati},
  {Clampin}, {Hartig}, {Zirm}, {R{\"o}ttgering}, {Venemans}, {Ardila},
  {Bartko}, {Broadhurst}, {Brown}, {Burrows}, {Cheng}, {Cross}, {De Breuck},
  {Feldman}, {Franx}, {Golimowski}, {Gronwall}, {Infante}, {Martel},
  {Menanteau}, {Meurer}, {Sirianni}, {Kimble}, {Krist}, {Sparks}, {Tran},
  {White}, \& {Zheng}}]{Miley2004}
{Miley}, G.~K., {et~al.} 2004, \nat, 427, 47

\bibitem[{{Miyazaki} {et~al.}(2002){Miyazaki}, {Komiyama}, {Sekiguchi},
  {Okamura}, {Doi}, {Furusawa}, {Hamabe}, {Imi}, {Kimura}, {Nakata}, {Okada},
  {Ouchi}, {Shimasaku}, {Yagi}, \& {Yasuda}}]{miyazaki2002}
{Miyazaki}, S., {et~al.} 2002, PASJ, 54, 833

\bibitem[{{Moore} {et~al.}(1996){Moore}, {Katz}, {Lake}, {Dressler}, \&
  {Oemler}}]{Moore1996}
{Moore}, B., {Katz}, N., {Lake}, G., {Dressler}, A., \& {Oemler}, A. 1996,
  \nat, 379, 613

\bibitem[{{Ouchi} {et~al.}(2004){Ouchi}, {Shimasaku}, {Okamura}, {Furusawa},
  {Kashikawa}, {Ota}, {Doi}, {Hamabe}, {Kimura}, {Komiyama}, {Miyazaki},
  {Miyazaki}, {Nakata}, {Sekiguchi}, {Yagi}, \& {Yasuda}}]{ouchi2004}
{Ouchi}, M., {et~al.} 2004, ApJ, 611, 660

\bibitem[{{Ouchi} {et~al.}(2005){Ouchi}, {Shimasaku}, {Akiyama}, {Sekiguchi},
  {Furusawa}, {Okamura}, {Kashikawa}, {Iye}, {Kodama}, {Saito}, {Sasaki},
  {Simpson}, {Takata}, {Yamada}, {Yamanoi}, {Yoshida}, \&
  {Yoshida}}]{Ouchi2005}
---. 2005, \apjl, 620, L1

\bibitem[{{Papovich} {et~al.}(2010){Papovich}, {Momcheva}, {Willmer},
  {Finkelstein}, {Finkelstein}, {Tran}, {Brodwin}, {Dunlop}, {Farrah}, {Khan},
  {Lotz}, {McCarthy}, {McLure}, {Rieke}, {Rudnick}, {Sivanandam}, {Pacaud}, \&
  {Pierre}}]{papovich2010}
{Papovich}, C., {et~al.} 2010, ApJ, 716, 1503

\bibitem[{{Pentericci} {et~al.}(1997){Pentericci}, {Roettgering}, {Miley},
  {Carilli}, \& {McCarthy}}]{pentericci1997}
{Pentericci}, L., {Roettgering}, H.~J.~A., {Miley}, G.~K., {Carilli}, C.~L., \&
  {McCarthy}, P. 1997, \aap, 326, 580

\bibitem[{{Pentericci} {et~al.}(2000){Pentericci}, {Van Reeven}, {Carilli},
  {R{\"o}ttgering}, \& {Miley}}]{Pentericci2000}
{Pentericci}, L., {Van Reeven}, W., {Carilli}, C.~L., {R{\"o}ttgering},
  H.~J.~A., \& {Miley}, G.~K. 2000, \aaps, 145, 121

\bibitem[{{Rocca-Volmerange} {et~al.}(2004){Rocca-Volmerange}, {Le Borgne}, {De
  Breuck}, {Fioc}, \& {Moy}}]{Rocca-Volmerange2004}
{Rocca-Volmerange}, B., {Le Borgne}, D., {De Breuck}, C., {Fioc}, M., \& {Moy},
  E. 2004, \aap, 415, 931

\bibitem[{{Rodighiero} {et~al.}(2011){Rodighiero}, {Daddi}, {Baronchelli},
  {Cimatti}, {Renzini}, {Aussel}, {Popesso}, {Lutz}, {Andreani}, {Berta},
  {Cava}, {Elbaz}, {Feltre}, {Fontana}, {F{\"o}rster Schreiber},
  {Franceschini}, {Genzel}, {Grazian}, {Gruppioni}, {Ilbert}, {Le Floch},
  {Magdis}, {Magliocchetti}, {Magnelli}, {Maiolino}, {McCracken}, {Nordon},
  {Poglitsch}, {Santini}, {Pozzi}, {Riguccini}, {Tacconi}, {Wuyts}, \&
  {Zamorani}}]{Rodighiero2011}
{Rodighiero}, G., {et~al.} 2011, \apjl, 739, L40

\bibitem[{{Salpeter}(1955)}]{salpeter1955}
{Salpeter}, E.~E. 1955, \apj, 121, 161

\bibitem[{{Schlegel} {et~al.}(1998){Schlegel}, {Finkbeiner}, \&
  {Davis}}]{schlegel1998}
{Schlegel}, D.~J., {Finkbeiner}, D.~P., \& {Davis}, M. 1998, ApJ, 500, 525

\bibitem[{{Seymour} {et~al.}(2007){Seymour}, {Stern}, {De Breuck}, {Vernet},
  {Rettura}, {Dickinson}, {Dey}, {Eisenhardt}, {Fosbury}, {Lacy}, {McCarthy},
  {Miley}, {Rocca-Volmerange}, {R{\"o}ttgering}, {Stanford}, {Teplitz}, {van
  Breugel}, \& {Zirm}}]{Seymour2007}
{Seymour}, N., {et~al.} 2007, \apjs, 171, 353

\bibitem[{{Sobral} {et~al.}(2012){Sobral}, {Best}, {Matsuda}, {Smail}, {Geach},
  \& {Cirasuolo}}]{sobral2012a}
{Sobral}, D., {Best}, P.~N., {Matsuda}, Y., {Smail}, I., {Geach}, J.~E., \&
  {Cirasuolo}, M. 2012, \mnras, 420, 1926

\bibitem[{{Steidel} {et~al.}(1998){Steidel}, {Adelberger}, {Dickinson},
  {Giavalisco}, {Pettini}, \& {Kellogg}}]{steidel1998}
{Steidel}, C.~C., {Adelberger}, K.~L., {Dickinson}, M., {Giavalisco}, M.,
  {Pettini}, M., \& {Kellogg}, M. 1998, ApJ, 492, 428

\bibitem[{{Suzuki} {et~al.}(2008){Suzuki}, {Tokoku}, {Ichikawa}, {Uchimoto},
  {Konishi}, {Yoshikawa}, {Tanaka}, {Yamada}, {Omata}, \&
  {Nishimura}}]{suzuki2008}
{Suzuki}, R., {et~al.} 2008, PASJ, 60, 1347

\bibitem[{{Tadaki} {et~al.}(2011){Tadaki}, {Kodama}, {Koyama}, {Hayashi},
  {Tanaka}, \& {Tokoku}}]{tadaki2011}
{Tadaki}, K.-I., {Kodama}, T., {Koyama}, Y., {Hayashi}, M., {Tanaka}, I., \&
  {Tokoku}, C. 2011, \pasj, 63, 437

\bibitem[{{Tanaka} {et~al.}(2011){Tanaka}, {Breuck}, {Kurk}, {Taniguchi},
  {Kodama}, {Matsuda}, {Packham}, {Zirm}, {Kajisawa}, {Ichikawa}, {Seymour},
  {Stern}, {Stockton}, {Venemans}, \& {Vernet}}]{tanaka.I2011}
{Tanaka}, I., {et~al.} 2011, \pasj, 63, 415

\bibitem[{{Tanaka} {et~al.}(2005){Tanaka}, {Kodama}, {Arimoto}, {Okamura},
  {Umetsu}, {Shimasaku}, {Tanaka}, \& {Yamada}}]{tanaka2005}
{Tanaka}, M., {Kodama}, T., {Arimoto}, N., {Okamura}, S., {Umetsu}, K.,
  {Shimasaku}, K., {Tanaka}, I., \& {Yamada}, T. 2005, MNRAS, 362, 268

\bibitem[{{Tran} {et~al.}(2010){Tran}, {Papovich}, {Saintonge}, {Brodwin},
  {Dunlop}, {Farrah}, {Finkelstein}, {Finkelstein}, {Lotz}, {McLure},
  {Momcheva}, \& {Willmer}}]{tran2010}
{Tran}, K., {et~al.} 2010, ApJL, 719, L126

\bibitem[{{Ueda} {et~al.}(2003){Ueda}, {Akiyama}, {Ohta}, \&
  {Miyaji}}]{ueda2003}
{Ueda}, Y., {Akiyama}, M., {Ohta}, K., \& {Miyaji}, T. 2003, ApJ, 598, 886

\bibitem[{{Venemans} {et~al.}(2005){Venemans}, {R{\"o}ttgering}, {Miley},
  {Kurk}, {De Breuck}, {Overzier}, {van Breugel}, {Carilli}, {Ford}, {Heckman},
  {Pentericci}, \& {McCarthy}}]{Venemans2005}
{Venemans}, B.~P., {et~al.} 2005, \aap, 431, 793

\bibitem[{{Venemans} {et~al.}(2007){Venemans}, {R{\"o}ttgering}, {Miley}, {van
  Breugel}, {de Breuck}, {Kurk}, {Pentericci}, {Stanford}, {Overzier}, {Croft},
  \& {Ford}}]{Venemans2007}
---. 2007, \aap, 461, 823

\bibitem[{{Villar-Mart{\'{\i}}n} {et~al.}(2007){Villar-Mart{\'{\i}}n},
  {S{\'a}nchez}, {Humphrey}, {Dijkstra}, {di Serego Alighieri}, {De Breuck}, \&
  {Gonz{\'a}lez Delgado}}]{VillarMartn2007}
{Villar-Mart{\'{\i}}n}, M., {S{\'a}nchez}, S.~F., {Humphrey}, A., {Dijkstra},
  M., {di Serego Alighieri}, S., {De Breuck}, C., \& {Gonz{\'a}lez Delgado}, R.
  2007, \mnras, 378, 416

\end{thebibliography}
\end{document}